# Chaos Phase Induced Mass-producible Monolayer Two-dimensional Material

Jiahui Wang, [1,4][†] Pengfei Yang,[1][†] Jing Jiang,[1] Yi Yan,[1] Jian-ping Ma,[2] Song Yang,[2] Luo Yang,[4] Qi-Kui Liu,[2*] Yin Chen[1,3*]

[1] Cent S Univ, Coll Chem & Chem Engn, Changsha 410083, Hunan, China

[2] Shandong Normal Univ, Coll Chem Chem Engn & Mat Sci, Collaborative Innovation Centre of Functionalized Probes for Chemical Imaging in Universities of Shandong, Key Laboratory of Molecular and Nano Probes, Ministry of Education, Jinan 250014, Shandong, China

[3] Key Laboratory of Hunan Province for Water Environment and Agriculture Product Safety, Changsha 410083, Hunan, China

[4] Xiangtan Univ, Coll Chem, Xiangtan 411105, Hunan, China

*Correspondence to: chenyin@iccas.ac.cn, qikuiliu2004@163.com.

† These authors contribute equally.

**Abstract**

Crystal phase is well-studied and presents a periodical atom arrangement in three-dimensions lattice, but the "amorphous phase" is poorly understood. Here, by starting from cage-like bicyclocalix[2]arene[2]triazines building block, a brand-new 2D MOF is constructed with extremely weak interlaminar interaction existing between two adjacent 2D-crystal layer. Inter-layer slip happens under external disturbance and leads to the loss of periodicity at one dimension in the crystal lattice, resulting in an interim phase between the crystal and amorphous phase - the chaos phase, non-periodical in microscopic scale but orderly in mesoscopic scale. This chaos phase 2D-MOF is a disordered self-assembly of black-phosphorus like 3D-layer, which has excellent mechanical-strength and a thickness of 1.15 nm. The bulky 2D-MOF material is readily to be exfoliated into monolayer nanosheets in gram-scale with unprecedented evenness and homogeneity, as well as previously unattained lateral size (>10 μm), which present the first mass-producible monolayer 2D material and can form wafer-scale film on substrate.

Solid materials are generally separated into crystal phase and amorphous phase depending on the periodicity of atom arrangement in the three dimension lattice.[1] Philosophically speaking, there must be interim states between these two extremes, but this grey area is always ignored. The discovery of quasicrystal by Dan Shechtman disclose the ordered but not periodic structure in metal alloys and open a window to this grey area.[2,3] Recently, two-dimension (2D) materials have been making a big splash and experiencing a boom,[4-8] which have atomic periodicity in two dimension and present a new crystal phase - 2D crystal.[9,10] By deliberately destruction of the periodicity at one dimension in the bulky 2D materials, a possibly interim solid state can be obtained in the solid material because of the partial loss of the atomic periodicity in crystal lattice (Fig.1).

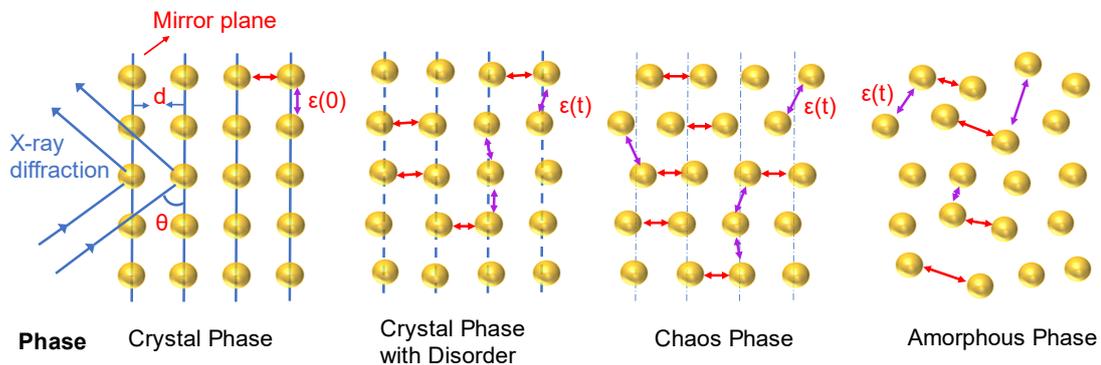

**Fig.1.** The schematic illustration of the degeneration from crystal pahse to amouphous phase, as well as the relationship between the atomic periodicity and the material phase.

Generally speaking, only weak Van da Waals interaction exist between adjacent 2D material layers.[11-13] However, millions of Van da Waals interaction pairs can greatly enhance the interaction force and periodicity of two adjacent layers,[14] which presents the biggest resistance for the production of monolayer 2D materials, and current bulky 2D materials always exist as crystal phase material. Elimination of the interlaminar interaction is the key point to achieve the proposed interim state, which also can make the mass-production and commercial applications of monolayer 2D materials possible.[15-19]

Unlike the graphene or transition metal dichalcogenides, MOF and COF can provide great design freedom for 2D materials.[20-25] We intend to design a 2D MOF with extremely low interlaminar interaction to test this hypothesis. As a rising member in 2D materials, MOF nanosheets possess large surface and highly accessible active sites on the surface, exhibiting unique performances in many different fields, such as molecular sieving, sensing, and catalysis.[26-31] Although we share a MOFs library with more than 20, 000 kinds, and the figure keeps growing quickly, the reported 2D MOF candidates suitable for the preparation of MOF nanosheets are very limited. The rational synthesis of 2D MOF with tailored structures and properties is not easy at all.[31-37]

An in-depth investigation indicates that the reported MOF nanosheets can be divided into two categories: One category has hydrocarbon group sandwiched layers, only weak Van der Waal's interactions exist between the adjacent layers; [31,37-40] The other category has steric porous layers, the smaller interlayer contact area can further decrease the interlaminar interaction.[41-44] Impressively, all these 2D MOFs have thick 3D-layered structure, however, the interlaminar interaction looks still strong as they still present as stable crystal phase material and the corresponding nanosheets can't be produced in large amount.

To exfoliate the bulky crystallized material into nanosheet, it's essentially a mechanics problem. When a bulky 2D MOF crystal is immersed in the solvent, the layers are subjected to propulsive force resulting from the molecule collision. The collision momentum component parallel to the MOF layer plus the viscous force equals the shear force ($F_s$) separating the layers. The component vertical to the MOF layers tends to tear the layers into smaller pieces. $F_s$ is evaluated by the following equation, which can be greatly enhanced along the direction of the driving force.[45] (detailed information in SI):

$$F_s = 2g \cdot l \cdot d \cdot h \cdot (1 + \Delta \cdot w) \int_{-90}^{90} \frac{dN_\theta}{N} \cos^2\theta \, v_s^2 + \eta \cdot w \cdot l \frac{dv_s}{dw} \cos\theta_v$$

Inter-layer interaction ($F_i$) is the interaction between adjacent layers, which correlates with the strength and density in unit area of the inter-layer pairwise interactions. $F_i$ competes with $F_s$ to avoid exfoliation and equals to:

$$F_i = l \cdot w \cdot [n_1 \frac{-6C_1}{r_1^7} cos\theta_1 + n_2 \frac{-6C_2}{r_2^7} cos\theta_2 + \cdots ]$$

Obviously, the inter-layer slip and layers separation happens only when $F_s > F_i$. Therefore, three key factors need to be considered in the preparation of monolayer nanosheet. First, the bulky crystal should have a layered structure and weak interlaminar interaction to reach the key $F_s > F_i$ criterion; Second, $F_s$ is nearly proportional to the layer thickness ($h$). Bigger $h$ is advantage, and porous structure also brings bigger effective force area and the resulting $F_s$; Last but not least, the layer should be mechanically strong to avoid damage caused by the propulsive force vertical to the layer plane. Clearly, in porous structure, this propulsive force and the damage can be effectively reduced. The lateral size and layer number of the exfoliated nanosheets are the results of the combination of these three factors.

As most of the current MOFs are constructed by 2D planar building blocks, it's very difficult, if not impossible, to design and synthesize the expected 2D MOF with extremely low interlaminar interaction and 3D-layered structure from 2D building blocks. We noticed that rigid 3D symmetric molecules tended to self-assemble into 3D-layered structures.[46-50] Bicyclocalix[2]arene[2]triazines, a cage-like $D_3h$ symmetric molecule we synthesized before, also self-assembled into a layered porous structure.[51] This molecule is lack of intermolecular-interaction sites and can adapt to different linkage groups, which most likely to afford the 2D-MOF with demanded structure. In addition, heterocalixaromatics not only have powerful recognition properties and functions, but are also rich in structures and conformations.[52,53] The cage-like host molecules are even more intriguing for their enclosed cavities and rigid structure, which can trap and stabilize specific guest molecules, work as molecular flask and endow intrinsic porosity.[54-57]

By using bicyclocalix[2]arene[2]triazines tricarboxylic acid (BCTA) as the building block and Mn cluster as the node, a porous 2D-MOF, CSUMOF-1, was prepared with extremely weak interlaminar interaction, high mechanical strength and 1.15 nm thick 3D-layer, which perfectly match our design requirements. As expected, inter-layer slip in the crystal lattice and loss of atomic periodicity in one dimension was observed. But macroscopically, CSUMOF-1 "crystal" still has a stability right to the point and presents an interim phase between the crystal and amorphous phase. CSUMOF-1 is readily to be exfoliated into monolayer nanosheets in gram-scale by simple ultrasonic exfoliation, which have a lateral size up to tens of micrometres, as well as unprecedent evenness and homogeneity.

BCTA tends to self-assemble into layered hydrogen-bonded organic frameworks (Fig. S2), which is easily weathered in the air (5 $m^2$/g BET surface area). The MOF crystals were easily synthesized from BCTA and $MnCl_2$ *via* solvothermal method. CSUMOF-1 was prepared in dimethylformamide (DMF) with a yield of 70%. In tetrahydrofuran, CSUMOF-2 was crystalized with a yield of 77% (Fig. S3-5).

In the begining, we obtained a very stable three-dimensional framework, CSUMOF-2, which had an expected layered structure containing one-dimensional rectangle channel with a size of 1.2 nm * 1.5 nm (Fig. 2A, S6, 7). The space in the channel was partly filled by the coordinated THF molecules. Single CSUMOF-2 layer is constituted by parallel Zigzag-arranged BCTA chains with a thickness of 0.6 nm, and bound by the [–Mn-Cl-Mn-O-]$_n$ inorganic chains to form solid 3D frameworks. CSUMOF-2 was stable up to 460 °C (Fig S8), with a BET surface area of 370 m$^2$/g (N$_2$, 77K) and $CO_2$ absorption of 21 cm$^3$/g (273K, Fig. 3). The experimental pore size distribution coincide with the expected values (Fig S9). Powder XRD pattern of CSUMOF-2 fits the simulated one very well (Figure S10), two main peaks at 8° and 11° are attributed to the diffraction of 002 and 020 crystal planes, both located on the rotatable benzene rings (Fig S11). After heating at 100 °C, peak shift and drastic intensity decrease were observed in the diffraction pattern, and subsequent exposure to THF only recovered the positions of the peaks; Soaking in solvents other than THF only caused the decrease of diffraction intensity. Similar results also have been observed in Cu(OPTz).[58] In CSUMOF-2, the channels can provide free space for the dynamic motion of the organic linkers (around one angstrom).[59-61] In the other hand, the coordination environment change of manganese may cause the atom position offset in the crystal lattice. Both these two factors lead to the disorder of the diffraction planes and degradation of the diffraction pattern. Generally speaking, CSUMOF-2 owes a strong frameworks with a certain flexibility .

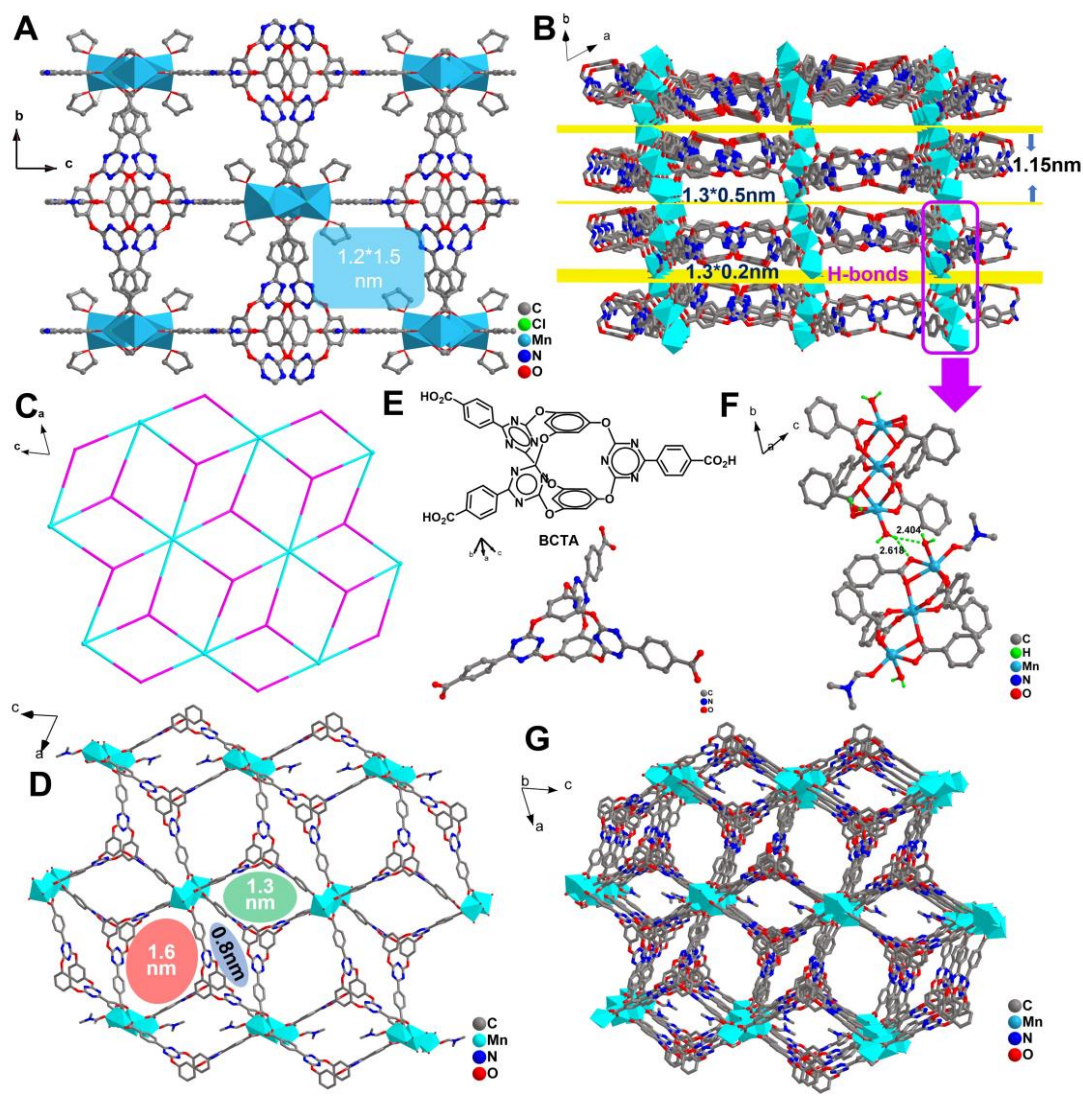

**Fig. 2.** Structure of CSUMOFs. (A) The presentation of one layer in CSUMOF-2. (B) Architecture of the layered structure of CSUMOF-1, with hydrogen-bond sites and channels between the layers. (C) The 3,6-connected net nodes topological structure for single CSUMOF-1 layer. (D) The presentation of one layer in CSUMOF-1 and the three kinds of pores with different sizes. (E) Molecular structure of BCTA. (F) The interlayer O–H···O hydrogen bonds between two adjacent different [Mn$_3$(O$_2$C)$_6$] clusters. (F) The ellipsoid channels formed by the packed CSUMOF-1 layers.

When DMF was used as the solvent, the Mn cluster bridge could be disconnected into [Mn$_3$(O$_2$C)$_6$] clusters. CSUMOF-1 has a separated layered structure, with a layer thickness as 1.15 nm, which nearly is the doubled height of BCTA (Fig. 2B). In the layer, there are two kinds of [Mn$_3$(O$_2$C)$_6$] clusters ([Mn$_3$(O$_2$C)$_6$]·4H$_2$O and [Mn$_3$(O$_2$C)$_6$]·2H$_2$O·2DMF, arranged into two alternatively rows in the layer plane, Fig. S12), which are connected by trigeminal BCTA ligands to form a 2D network with 3,6-connected net nodes (Schläfli symbol {4^3}2{4^6;6^6;8^3}, Fig. 2C). Topologically, this network is very rigid with high mechanical strength, extending in the *ac* plane with three types of regular pores (the pore sizes are 0.8nm, 1.3nm and 1.5nm respectively, Fig. 2D). Along [0, -1, 1] axis, one single layer looks like square-wave, similar to black-phosphorus layer in some extent. The net-like layers stacked in ABAB fashion along the crystallographic *b* axis *via* weak inter-layer O–H···O hydrogen bonds [$d_{O-H\cdots O}$ = 2.618 and 2.404 Å] between these two different [$Mn_3(O_2C)_6$] clusters (the sole interaction between the layers, Fig. 2F, S12, S13), resulting in a 3D neutral supramolecular framework with three ellipsoid channels (Fig. 2G). There are two types of regular rectangle channels existing between the square-wave layers for a half-phase difference (the sizes are 1.3*0.5 nm and 1.3*0.2 nm respectively, Fig. S14). All these channels vertical and parallel to the layer surface have significantly decreased the real contact area and the hydrogen-bond number in unit area between two conjunction layers. In brief, extremely low interlaminar interaction can be expected, which perfectly meet our design requirements.

CSUMOF-1 looks very "unstable". A powder XRD pattern with poor quality could be recorded only with freshly-prepared samples, but fitted the simulated one (Fig. 3A). After heating, drying or solvent exchange, CSUMOF-1 became "amorphous" (Fig. 3B), without any change of the crystal appearance (Fig. S15). These results were very confusing but not unexpected, which can be attributed to the intrinsic macroscopic properties of the proposed interim phase.

CSUMOF-1 has a BET surface area of 210 $m^2/g$ and $CO_2$ absorption of 41 $cm^3/g$ after activation at 150 °C under vacuum, and good thermal stability up to 440 °C (Fig. S8). Similar to CSUMOF-2, CSUMOF-1 has a typical type I isotherms in $N_2$ adsorption-desorption. The adsorption was apparent in the low-pressure region (P/P$_0$ < 0.05), and the desorption of $N_2$ was reversible and revealed hysteresis. The difference is that CSUMOF-1 has nearly doubled $CO_2$ absorption with halved BET surface area. Heterocalixaromatics can bind $CO_2$ by its nitrogen-rich cavities.[48] Obviously, the remarkable $CO_2$ absorption of CSUMOF-1 can be attributed to its fully exposed calixarene cavities in the lattice. Experimental results disclosed that the blocked channels and calixarene cavities in CSUMOF-1 could be reopened at high pressure. At 273 K, when the pressure reached 60 atm (maximum 200 atm), $N_2$ absorption of CSUMOF-1 increased from 2.8 $cm^3/g$ to 7.0 $cm^3/g$ (the interaction between $N_2$ molecules and the frameworks is very weak). Meanwhile, the $CO_2$ absorption (performed on the same sample) increased from 18 $cm^3/g$ to 56 $cm^3/g$ at 38 atm (Fig. 3E). Both desorption revealed hysteresis, but were reversible at reduced pressure, strongly supporting the occurrence of inter-layer slip and high structural-stability of the layer. [62,63]

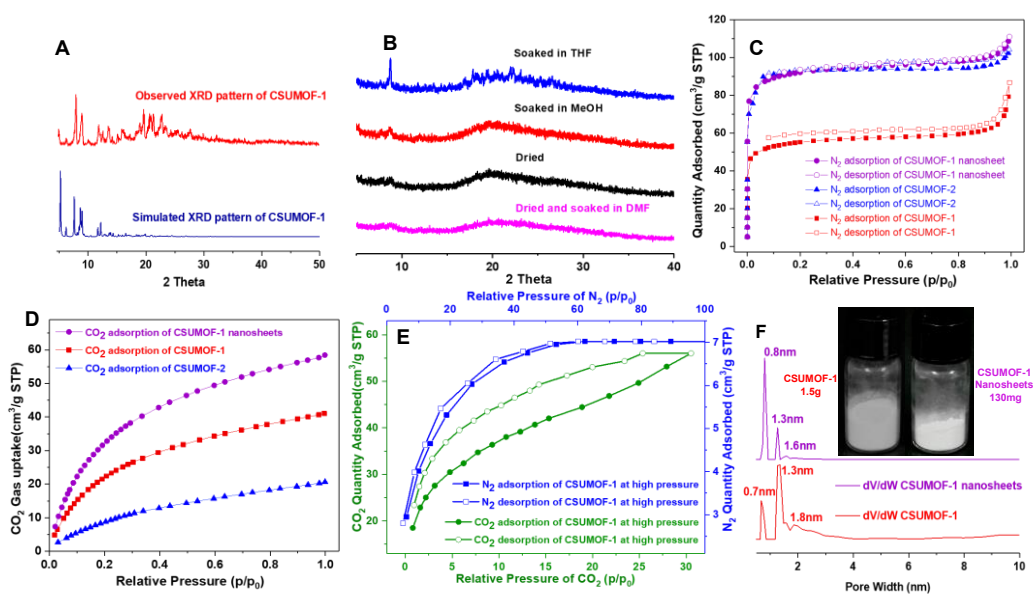

**Fig. 3.** Powder XRD and BET tests. (A) XRD powder pattern of fresh prepared CSUMOF-1 crystal and the simulated one. (B) XRD powder pattern of CSUMOF-1 crystal after drying or soaked in other solvent. (C) $N_2$ adsorption-desorption isotherms of CSUMOF-1(red), CSUMOF-2 (blue), and CSUMOF-1 nanosheets (purple) at 77 K and pressures up to 1 bar. (D) $CO_2$ adsorption isotherms of CSUMOF-1(red), CSUMOF-2 (blue), and CSUMOF-1 nanosheets (purple) at 273 K and pressures up to 1 bar. (E) $CO_2$ (273K) and $N_2$ (298K) adsorption-desorption isotherms of CSUMOF-1 at high pressure. (F) Narrow

pore-size distribution of CSUMOF-1 and CSUMOF-1 nanosheet. Inset, photographs comparing the CSUMOF-1 crystals and CSUMOF-1 nanosheets.

As observed by SEM images, CSUMOF-1 crystal has very clear and smooth layered structure (Fig. 4), which is readily to be exfoliated into nanosheets by routine ultrasonication treatment in different solvents (Fig. 4A). The colloid suspension remained stable at room temperature for several months, and significant Tyndall effect was observed (Fig. 4B inset, S16). Lots of ultrathin nanosheets with big lateral size and high evenness were observed in the suspension by TEM (Fig. 4, S17, S18). Only nano-particles were found in the sonicated suspension of CSUMOF-2 (Fig. S19, S20).

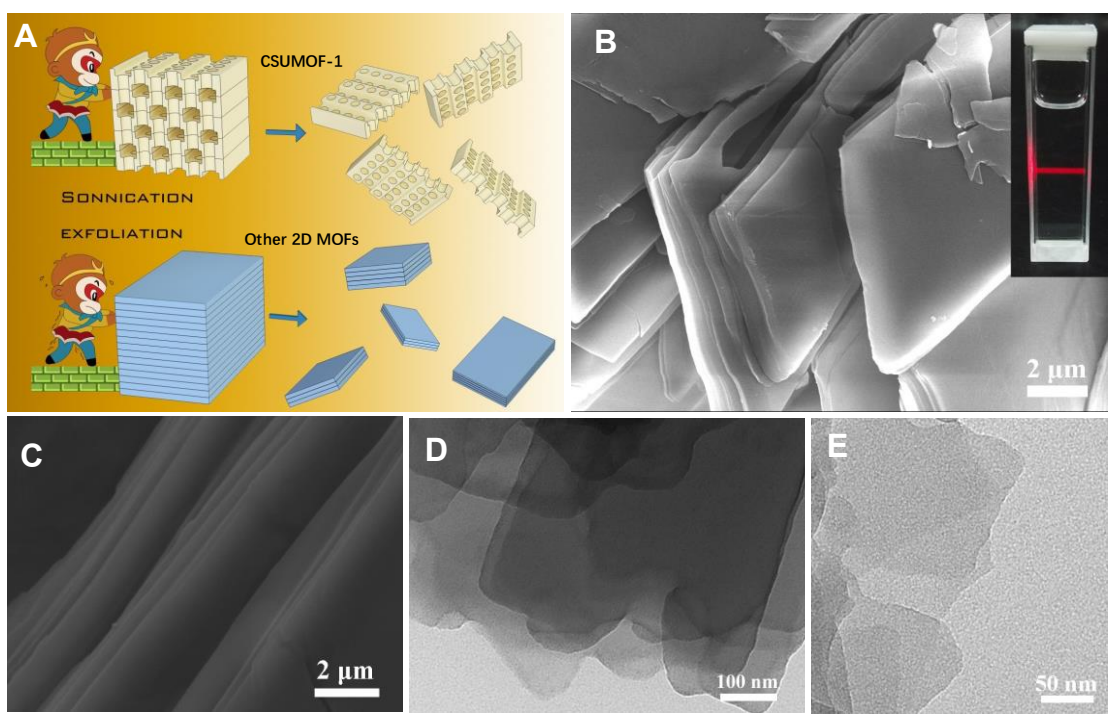

**Fig. 4.** Exfoliation of CSUMOF-1 and the electron microscopy images (A) Schematic illustration of the exfoliation of CSUMOF-1. (B) Scanning electron microscopy (SEM) image of CSUMOF-1 crystal as-synthesized. Inset, Tyndall effect of the nanosheets colloid suspension. (C) The clear and smooth layered architecture of CSUMOF-1. (D) Transmission electron microscopy (TEM) image of CSUMOF-1 nanosheets. (E) High-magnification TEM image of CSUMOF-1 nanosheets.

Tapping-mode atomic force microscopy (AFM) was applied to further identify these nanosheets.

Plenty of rigid and high flatness nanosheets were found in the high concentration nanosheets suspension (~ 0.1- 0.2 mg/ml, Figure 5, S21). The nanosheets have a lateral size range from several to more than ten micrometers. With a 1-5 μg/ml nanosheets suspension, the nanosheets formed a large-size wafer on the substrate with a thickness around 1-2 nm (Fig. 5A). Dispersed nanosheets was observed after further dilution. Fig. 5C shows an enlarged area of the nanosheet. The height profile reveals that the nanosheet is extremely flat, with a smooth terrace height as 1.12nm, which fits the layer thickness value in the single-crystal structure. As demonstrated by the AFM analyses on 43 different sites (Fig. 5E, S22), more than 90% of them had a thickness of 1.1±0.2nm, and confirmed the nanosheets as monolayer with high homogeneity.

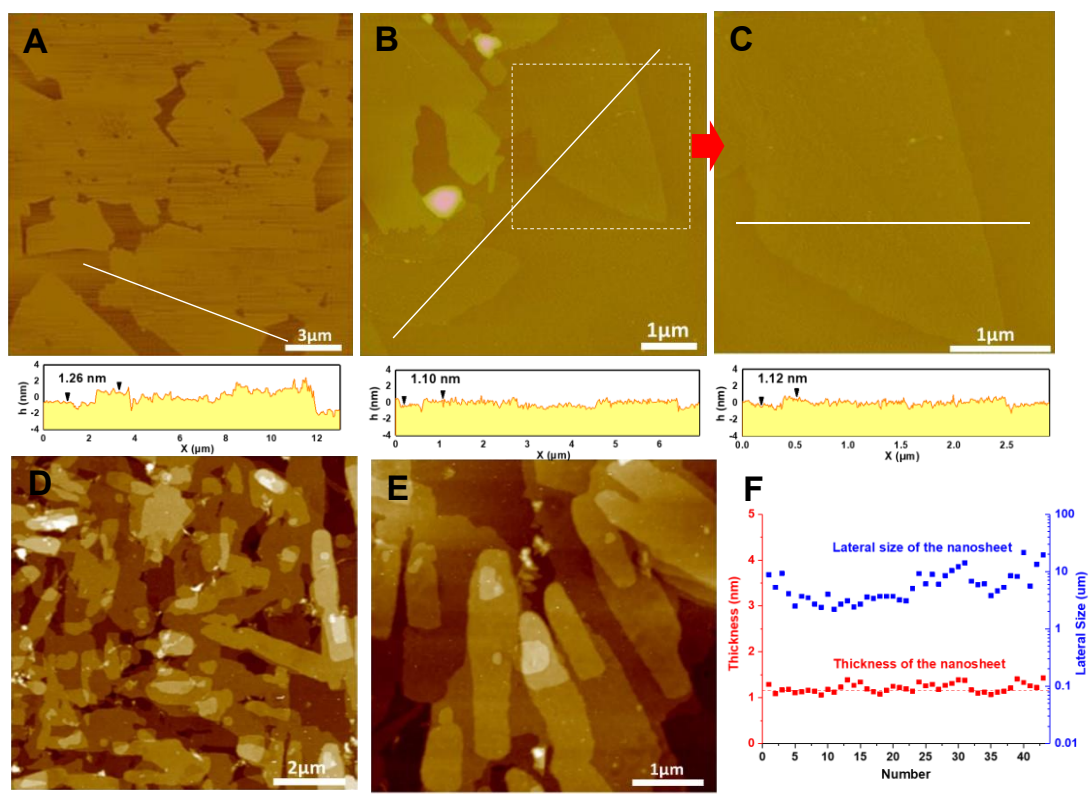

**Fig. 5**. AFM images of CSUMOF-1 nanosheets (a-c) with low concentration suspension on freshly cleaved mica surfaces. The height profile of the nanosheets along the white lines were marked below. (d-e) with high concentration suspension on freshly cleaved mica surfaces. (f) Thickness and lateral size distribution of exfoliated CSUMOF-1 nanosheets (horizontal line indicates the theoretical thickness of a single layered CSUMOF-1 nanosheets).

The infrared spectrum of nanosheets prepared in large-batch is perfectly coincide with that of fresh-prepared CSUMOF-1(Fig. S23), but the absorbance bands in the fingerprint region become stronger and clearer. CSUMOF-1 nanosheets has nearly doubled BET surface area (360 $m^2/g$) and $CO_2$ absorption (58 $cm^3/g$) (Fig. 3C, 3D, performed with 130 mg nanosheets). More importantly, compare to bulky CSUMOF-1, the nanosheets has very sharp and clear pore width distribution (Fig. 3F), which is identical to the theoretical values (Fig. 2D). Apparently, the irregular pores in bulky CSUMOF-1 caused by the inter-layer slip has disappeared after complete separation, and no hysteresis could be found anymore in the $N_2$ desorption isotherm of nanosheet (Fig. S24). It also can be concluded that the inter-layer slip in CSUMOF-1 is around several angstroms. These results unambiguously confirmed the chemical structure and macroscopic homogeneity of the as-prepared nanosheets,

The rigid and smooth appearance indicates that the nanosheets should have high mechanical strength. The extremely weak interlaminar interaction and nanometer layer thickness observed in the crystal structure provide sufficient and necessary conditions to satisfy the key $F_s > F_i$ criterion. Thus, CSUMOF-1 was exfoliated into monolayer nanosheets in gram-scale with big lateral size. All these experimental results have strongly supported our theoretical hypothesis on the interim phase and monolayer 2D materials production. The cheap, easy and scalable preparation of monolayer CSUMOF-1 nanosheets also make its commercial applications possible.

According to classic XRD crystallography, CSUMOF-1 should be classified as "amorphous". However, the experiment results unambiguously indicated that CSUMOF-1 is a stable self-assembly of 2D-crystals with disorder along the dimension vertical to the 2D-crystal plane. In the crystal lattice, the atoms do chaos thermal motion in confined space, which leads to the loss of periodicity. If the initial distance between two adjacent atoms is set as ε(0), after time *t*, the distance can be described as $\varepsilon(t) = \varepsilon(0)e^{\lambda t}$ ($\lambda = \lim_{n\to\infty} \frac{1}{n} \sum_{n=0}^{n-1} ln \left|\frac{d\varepsilon(t)}{d\varepsilon(0)}\right|$, the Lyapunov exponent, Fig. 1, S25).[64,65] For an effective diffraction crystal plane with n atoms, the detection error will be amplified by $e^{(n-1)\lambda}$ times. In idea closely-packed crystal phase, the thermal motion is confined in picometers scale. Meanwhile, the lattice wave makes the atoms correlated in long range, $\varepsilon(t) \approx \varepsilon(0)$, λ→0, the error of the diffraction crystal planes can be omitted. In porous frameworks, such as CUSMOF-2, Cu(OPTz) etc., the free space in the crystal lattice encourages the dynamic motion of the linkers. $\varepsilon_{(max)}$ of the atoms in the linker will increase to angstrom scale as this dynamic motion is independent from the lattice wave. At present, the crystal periodicity generally is studied by X-Ray with a precision at sub-angstrom scale. The increased $\varepsilon_{(max)}$ leads to the significant increased λ value and error of the diffraction planes with rotatable linkers, as well as degenerated S/N ratio. Thus, broadening and weakening XRD peaks are observed as the result.

In CUSMOF-1, the adjacent layers are only bound by weak hydrogen bonds. Experimental results indicate that the movement of the atoms along the layer plane can increase to several angstroms due to the inter-layer slip. Given a 5 Å $\varepsilon_{(max)}$, $\lambda_{CUSMOF-1}$ will increase around 5 times comparing to $\lambda_{CUSMOF-2}$ (λ value is decided by $\lambda_{max} = \frac{1}{t_m-t_0}\sum_{k=1}^{m} ln \frac{\varepsilon(t_k)}{\varepsilon(t_{k-1})}$). The corresponding error will increase 50 times (more than $10^7$ times for 1 nm$^3$ confined space). Diffraction signal can't be detected due to a simultaneously decreased S/N ratio, and dried CUSMOF-1 presents as "amorphous" material. It should be noted that even nanosheets with few layers could hardly be detected according to Bragg's Law. However, if the measurement accuracy decreases to nanometer-scale (same scale as one CUSMOF-1 layer), λ value will decrease ten times and the error caused by the chaos can be omitted. Thus, the structural order is observed by corresponding characterization methods. We name this interim state as "Chaos phase", which is non-periodical in microscopic scale but order in mesoscopic scale.

**Reference:**

1   Marder, M. P. *Condensed matter physics*.   (John Wiley & Sons, 2010).
2   Shechtman, D., Blech, I., Gratias, D. & Cahn, J. W. Metallic phase with long-range orientational order and no translational symmetry. *Physical review letters* **53**, 1951 (1984).
3   Cahn, J. W., Shechtman, D. & Gratias, D. Indexing of icosahedral quasiperiodic crystals. *Journal of materials research* **1**, 13-26 (1986).
4   Novoselov, K. S. *et al.* Electric field effect in atomically thin carbon films. *Science* **306**, 666-669, doi:10.1126/science.1102896 (2004).
5   Novoselov, K. S. *et al.* Two-dimensional gas of massless Dirac fermions in graphene. *Nature* **438**, 197-200, doi:10.1038/nature04233 (2005).
6   Mannix, A. J., Kiraly, B., Hersam, M. C. & Guisinger, N. P. Synthesis and chemistry of elemental 2D materials. *Nature Reviews Chemistry* **1**, 0014, doi:10.1038/s41570-016-0014 (2017).
7   Tan, C. *et al.* Recent Advances in Ultrathin Two-Dimensional Nanomaterials. *Chemical Reviews* **117**, 6225-6331, doi:10.1021/acs.chemrev.6b00558 (2017).
8   Akinwande, D. *et al.* Graphene and two-dimensional materials for silicon technology. *Nature* **573**, 507-518, doi:10.1038/s41586-019-1573-9 (2019).
9   Ji, D. *et al.* Freestanding crystalline oxide perovskites down to the monolayer limit. *Nature* **570**, 87-+, doi:10.1038/s41586-019-1255-7 (2019).
10  Meng, S., Kong, T., Ma, W., Wang, H. & Zhang, H. 2D Crystal-Based Fibers: Status and Challenges. *Small* **15**, doi:10.1002/smll.201902691 (2019).
11  Rhodes, D., Chae, S. H., Ribeiro-Palau, R. & Hone, J. Disorder in van der Waals heterostructures of 2D materials. *Nature Materials* **18**, 541-549, doi:10.1038/s41563-019-0366-8 (2019).
12  Novoselov, K. S., Mishchenko, A., Carvalho, A. & Castro Neto, A. H. 2D materials and van der Waals heterostructures. *Science* **353**, aac9439 doi:10.1126/science.aac9439 (2016).
13  Liu, Y., Huang, Y. & Duan, X. F. Van der Waals integration before and beyond two-dimensional materials. *Nature* **567**, 323-333, doi:10.1038/s41586-019-1013-x (2019).
14  Lehn, J.-M. Supramolecular chemistry. *Science* **260**, 1762-1764, doi:10.1126/science.8511582 (1993).
15  Shim, J. *et al.* Controlled crack propagation for atomic precision handling of wafer-scale two-dimensional materials. *Science* **362**, 665-+, doi:10.1126/science.aat8126 (2018).
16  Zhong, Y. *et al.* Wafer-scale synthesis of monolayer two-dimensional porphyrin polymers for hybrid superlattices. *Science*, eaax9385, doi:10.1126/science.aax9385 (2019).

| 17 | Deng, Y. *et al.* Gate-tunable room-temperature ferromagnetism in two-dimensional Fe3GeTe2. *Nature* **563**, 94-+, doi:10.1038/s41586-018-0626-9 (2018). |
|---|---|
| 18 | Novoselov, K. S. Rapid progress in producing graphene. *Nature* **505**, 291-291, doi:10.1038/505291c (2014). |
| 19 | Sahoo, P. K., Memaran, S., Xin, Y., Balicas, L. & Gutierrez, H. R. One-pot growth of two-dimensional lateral heterostructures via sequential edge-epitaxy. *Nature* **553**, 63-+, doi:10.1038/nature25155 (2018). |
| 20 | Furukawa, H., Cordova, K. E., O'Keeffe, M. & Yaghi, O. M. The Chemistry and Applications of Metal-Organic Frameworks. *Science* **341**, 974-+, doi:10.1126/science.1230444 (2013). |
| 21 | Zhang, C., Wu, B.-H., Ma, M.-Q., Wang, Z. & Xu, Z.-K. Ultrathin metal/covalent-organic framework membranes towards ultimate separation. *Chemical Society Reviews* **48**, 3811-3841, doi:10.1039/c9cs00322c (2019). |
| 22 | Diercks, C. S. & Yaghi, O. M. The atom, the molecule, and the covalent organic framework. *Science* **355**, eaal1585, doi:10.1126/science.aal1585 (2017). |
| 23 | Colson, J. W. *et al.* Oriented 2D Covalent Organic Framework Thin Films on Single-Layer Graphene. **332**, 228-231, doi:10.1126/science.1202747 (2011). |
| 24 | Furukawa, S., Reboul, J., Diring, S., Sumida, K. & Kitagawa, S. Structuring of metal-organic frameworks at the mesoscopic/macroscopic scale. *Chemical Society Reviews* **43**, 5700-5734, doi:10.1039/c4cs00106k (2014). |
| 25 | Tanaka, D. *et al.* Rapid preparation of flexible porous coordination polymer nanocrystals with accelerated guest adsorption kinetics. *Nature Chemistry* **2**, 410-416, doi:10.1038/nchem.627 (2010). |
| 26 | Zhao, M. *et al.* Two-dimensional metal-organic framework nanosheets: synthesis and applications. *Chemical Society Reviews* **47**, 6267-6295, doi:10.1039/c8cs00268a (2018). |
| 27 | Guan, B. Y., Yu, X. Y., Wu, H. B. & Lou, X. W. Complex Nanostructures from Materials based on Metal-Organic Frameworks for Electrochemical Energy Storage and Conversion. *Advanced Materials* **29**, 1703614, doi:10.1002/adma.201703614 (2017). |
| 28 | Rodenas, T. *et al.* Metal–organic framework nanosheets in polymer composite materials for gas separation. *Nature Materials* **14**, 48, doi:10.1038/nmat4113 (2014). |
| 29 | Li, H.-C. *et al.* Metal-Organic Framework Templated Pd@PdO-Co3O4 Nanocubes as an Efficient Bifunctional Oxygen Electrocatalyst. *Advanced Energy Materials* **8**, 1702734, doi:10.1002/aenm.201702734 (2018). |
| 30 | Rui, K. *et al.* Hybrid 2D Dual-Metal-Organic Frameworks for Enhanced Water Oxidation Catalysis. *Advanced Functional Materials* **28**, doi:10.1002/adfm.201801554 (2018). |
| 31 | Peng, Y. *et al.* Metal-organic framework nanosheets as building blocks for molecular sieving membranes. *Science* **346**, 1356-1359, doi:10.1126/science.1254227 (2014). |
| 32 | Cai, X. K., Luo, Y. T., Liu, B. & Cheng, H. M. Preparation of 2D material dispersions and their applications. *Chemical Society Reviews* **47**, 6224-6266, doi:10.1039/c8cs00254a (2018). |
| 33 | Li, P.-Z., Maeda, Y. & Xu, Q. Top-down fabrication of crystalline metal-organic framework nanosheets. *Chemical Communications* **47**, 8436-8438, doi:10.1039/c1cc12510a (2011). |
| 34 | Abherve, A., Manas-Valero, S., Clemente-Leon, M. & Coronado, E. Graphene related magnetic materials: micromechanical exfoliation of 2D layered magnets based on bimetallic anilate complexes with inserted Fe-III(acac(2)-trien) (+) and Fe-III(sal(2)-trien) (+) molecules. *Chemical Science* **6**, 4665-4673, doi:10.1039/c5sc00957j (2015). |

35  Ding, Y. *et al.* Controlled Intercalation and Chemical Exfoliation of Layered Metal–Organic Frameworks Using a Chemically Labile Intercalating Agent. *Journal of the American Chemical Society* **139**, 9136-9139, doi:10.1021/jacs.7b04829 (2017).

36  Huang, J. *et al.* Electrochemical Exfoliation of Pillared-Layer Metal-Organic Framework to Boost the Oxygen Evolution Reaction. *Angewandte Chemie-International Edition* **57**, 4632-4636, doi:10.1002/anie.201801029 (2018).

37  Wang, X. *et al.* Reversed thermo-switchable molecular sieving membranes composed of two-dimensional metal-organic nanosheets for gas separation. *Nature Communications* **8**, 14460, doi:10.1038/ncomms14460 (2017).

38  Tan, J.-C., Saines, P. J., Bithell, E. G. & Cheetham, A. K. Hybrid Nanosheets of an Inorganic-Organic Framework Material: Facile Synthesis, Structure, and Elastic Properties. *Acs Nano* **6**, 615-621, doi:10.1021/nn204054k (2012).

39  Hermosa, C. *et al.* Mechanical and optical properties of ultralarge flakes of a metal-organic framework with molecular thickness. *Chemical Science* **6**, 2553-2558, doi:10.1039/c4sc03115f (2015).

40  Au, V. K.-M. *et al.* Stepwise Expansion of Layered Metal-Organic Frameworks for Nonstochastic Exfoliation into Porous Nanosheets. *Journal of the American Chemical Society* **141**, 53-57, doi:10.1021/jacs.8b09987 (2019).

41  Wang, Y. *et al.* Photosensitizer-Anchored 2D MOF Nanosheets as Highly Stable and Accessible Catalysts toward Artemisinin Production. *Advanced Science* **6**, 1802059, doi:10.1002/advs.201802059 (2019).

42  Zuo, Q. *et al.* Ultrathin Metal-Organic Framework Nanosheets with Ultrahigh Loading of Single Pt Atoms for Efficient Visible-Light-Driven Photocatalytic H2 Evolution. *Angewandte Chemie (International ed. in English)* **58**, 10198-10203, doi:10.1002/anie.201904058 (2019).

43  Xu, M. *et al.* Two-Dimensional Metal–Organic Framework Nanosheets as an Enzyme Inhibitor: Modulation of the α-Chymotrypsin Activity. *Journal of the American Chemical Society* **139**, 8312-8319, doi:10.1021/jacs.7b03450 (2017).

44  Zhan, G. *et al.* Fabrication of Ultrathin 2D Cu-BDC Nanosheets and the Derived Integrated MOF Nanocomposites. *Advanced Functional Materials* **29**, 1806720, doi:10.1002/adfm.201806720 (2019).

45  Israelachvili, J. N. *Intermolecular and surface forces*.  (Academic press, 2015).

46  Zhang, C. *et al.* A Porous Tricyclooxacalixarene Cage Based on Tetraphenylethylene. *Angewandte Chemie International Edition* **54**, 9244-9248, doi:10.1002/anie.201502912 (2015).

47  Chandrasekhar, P., Mukhopadhyay, A., Savitha, G. & Moorthy, J. N. Orthogonal self-assembly of a trigonal triptycene triacid: signaling of exfoliation of porous 2D metal-organic layers by fluorescence and selective CO2 capture by the hydrogen-bonded MOF. *Journal of Materials Chemistry A* **5**, 5402-5412, doi:10.1039/c6ta11110f (2017).

48  Wang, Z. *et al.* Networked Cages for Enhanced CO2 Capture and Sensing. *Advanced Science* **5**, 1800141, doi:10.1002/advs.201800141 (2018).

49  Kissel, P., Murray, D. J., Wulftange, W. J., Catalano, V. J. & King, B. T. A nanoporous two-dimensional polymer by single-crystal-to-single-crystal photopolymerization. *Nature Chemistry* **6**, 774-778, doi:10.1038/nchem.2008 (2014).

50  Kory, M. J. *et al.* Gram-scale synthesis of two-dimensional polymer crystals and their structure analysis by X-ray diffraction. *Nature Chemistry* **6**, 779-784, doi:10.1038/nchem.2007 (2014).

51	Chen, Y., Wang, J., Hai, X., Li, Q. & Jiang, J. Facial one-pot synthesis of D-3h symmetric bicyclocalix 2 arene 2 triazines and their layered comb self-assembly. *Journal of Saudi Chemical Society* **22**, 628-635, doi:10.1016/j.jscs.2017.11.003 (2018).

52	Wang, M.-X. Nitrogen and Oxygen Bridged Calixaromatics: Synthesis, Structure, Functionalization, and Molecular Recognition. *Accounts of Chemical Research* **45**, 182-195, doi:10.1021/ar200108c (2012).

53	Kim, D. S. & Sessler, J. L. Calix 4 pyrroles: versatile molecular containers with ion transport, recognition, and molecular switching functions. *Chemical Society Reviews* **44**, 532-546, doi:10.1039/c4cs00157e (2015).

54	Zhang, G. & Mastalerz, M. Organic cage compounds - from shape-persistency to function. *Chemical Society Reviews* **43**, 1934-1947, doi:10.1039/c3cs60358j (2014).

55	Mitra, T. *et al.* Molecular shape sorting using molecular organic cages. *Nature Chemistry* **5**, 276-281, doi:10.1038/nchem.1550 (2013).

56	Yoshizawa, M., Tamura, M. & Fujita, M. Diels-alder in aqueous molecular hosts: Unusual regioselectivity and efficient catalysis. *Science* **312**, 251-254, doi:10.1126/science.1124985 (2006).

57	Inokuma, Y. *et al.* X-ray analysis on the nanogram to microgram scale using porous complexes. *Nature* **495**, 461-+, doi:10.1038/nature11990 (2013).

58	Gu, C. *et al.* Design and control of gas diffusion process in a nanoporous soft crystal. *Science* **363**, 387-391, doi:10.1126/science.aar6833 (2019).

59	Horike, S. *et al.* Dynamic motion of building blocks in porous coordination polymers. *Angewandte Chemie-International Edition* **45**, 7226-7230, doi:10.1002/anie.200603196 (2006).

60	Gould, S. L., Tranchemontagne, D., Yaghi, O. M. & Garcia-Garibay, M. A. Amphidynamic character of crystalline MOF-5: Rotational dynamics of terephthalate phenylenes in a free-volume, sterically unhindered environment. *Journal of the American Chemical Society* **130**, 3246-+, doi:10.1021/ja077122c (2008).

61	Horike, S., Shimomura, S. & Kitagawa, S. Soft porous crystals. *Nature Chemistry* **1**, 695-704, doi:10.1038/nchem.444 (2009).

62	Kitaura, R., Seki, K., Akiyama, G. & Kitagawa, S. Porous coordination-polymer crystals with gated channels specific for supercritical gases. *Angewandte Chemie-International Edition* **42**, 428-431, doi:10.1002/anie.200390130 (2003).

63	Zhang, J.-P., Zhou, H.-L., Zhou, D.-D., Liao, P.-Q. & Chen, X.-M. Controlling flexibility of metal-organic frameworks. *National Science Review* **5**, 907-919, doi:10.1093/nsr/nwx127 (2018).

64	Williams, G. *Chaos theory tamed*. (CRC Press, 1997).

65	Chávez-Carlos, J. *et al.* Quantum and Classical Lyapunov Exponents in Atom-Field Interaction Systems. *Physical Review Letters* **122**, 024101, doi:10.1103/PhysRevLett.122.024101 (2019).

# Supplementary information for

**Chaos Phase Induced Mass-producible Monolayer Two-dimensional Material**

Jiahui Wang, [1,3][†] Pengfei Yang,[1][†] Jing Jiang,[1] Yi Yan,[1] Jianping Ma,[2] Song Yang,[2] Luo Yang,[3] QiKui Liu,[2][*] Yin Chen[1,4][*]

Correspondence to: chenyin@iccas.ac.cn (Y. C.) qikuiliu2004@163.com (Q. L.)

**The derivation of the equations:**

**Derivation of the formula for $F_s$:**

$F_s$ is composed by two parts, the impact force caused by the collision of solvent molecules ($F_m$) and the viscous force along the layer face ($F_v$). As illustrated in Fig. S1, according to the theory of Momentum, $F_m$ equal to the impulse acting on unit area of MOF layer in one second. The collision between solvent molecules and the MOF layers can be treated as elastic collision, it can be concluded that $F_m = 2 \cdot m \cdot v$. Due to the random movement of the solvent molecules, the weight of the solvent molecule acting on the MOF layer in one second equal to $d \cdot V \cdot cos\theta$. $d$ is the density of the solvent, V is the volume of solvent and equal to $l \cdot h \cdot v_s$. Only the collision momentum component parallel to the MOF layers contribute to $F_m$. $F_v$ is origin from the intermolecular interactions between the solvent molecules and the solid surface, which can be derived by Newton's law of viscous flow, $F = \eta S \frac{dv}{dx}$. In exfoliation, the fluid velocity decreased along the $w$ direction.

The equation for $F_s$ can be derived as:

$$F_s = 2g \cdot l \cdot d \cdot h \cdot (1 + \Delta \cdot w) \int_{-90}^{90} \frac{dN_\theta}{N} \cos^2\theta \, v_s^2 + \eta \cdot w \cdot l \frac{dv_s}{dw} cos\theta_v$$

$g$ = the ratio of gravity to mass

$l$ = Lateral size of the MOF layer along the ultrasonic wave (m)

$h$ = Thickness of the MOF layer (m)

$d$ = Density of the solvent (kg/m³)

$N$ = The total number of the collided solvent molecules

$\theta$ = The angle (º) between the solvent molecule motion direction and the MOF layer plane (as shown in Figure S1)

$\Delta$ = Area increment coefficient. For materials with uneven surface or big pores, the solvent molecules collide on the vertical part of the surface or the pore wall, which also contribute to the propulsive force, $n$ presents the area increment coefficient along the layer surface in unit width.

$v_s$ = The increment of solvent molecule velocity relative to the solid material powered by ultrasonic treatment (m/s), $v_s$ is proportional to the square root of power input ($P_i$), which is inversely proportional to $r^3$ ($r$ = the distance to the driving force source). In unit volume V, $P_i \times t = \frac{1}{2} g \times V \times d \times v_s^2$. Along the direction of *w*, $v_s$ decreased proportional to $r^{3/2}$, here $v_s$ can be treated as constant because *w* is in micrometer scale.

$\eta$ = The viscous coefficient (N·s/m$^2$). In most cases, solvent used in exfoliation has low $\eta$ value. $F_v$ only takes up a small proportion in $F_s$ for porous material.

$\theta_v$ = The angle (º) between the direction of V$_s$ and the MOF layer plane (as shown in Figure S1)

**Derivation of the formula for $F_i$:**

The $F_i$ can be calcualted based on the single crystal data, in one crystal cell, the type and quantity of the pairwise interactions between two adjacent layers can be listed, the equation for F$_i$ can be derived as:

$$F_i = l \cdot w \cdot [n_1 \frac{-6C_1}{r_1^7} cos\theta_1 + n_2 \frac{-6C_2}{r_2^7} cos\theta_2 + \cdots ]$$

*l* = the length of the MOF layer (nm)

*w* = the width of the MOF layer (nm)

$n_1$ = the number of interacted atom pairs per square nanometer in type 1 (nm$^{-2}$)

$C_1$ = London constant

$r_1$ = the distance between the interacted pairwise atoms in Type 1 (m)

$\theta_1$, $\theta_2$ = the angle between the interaction direction and the MOF layer plane

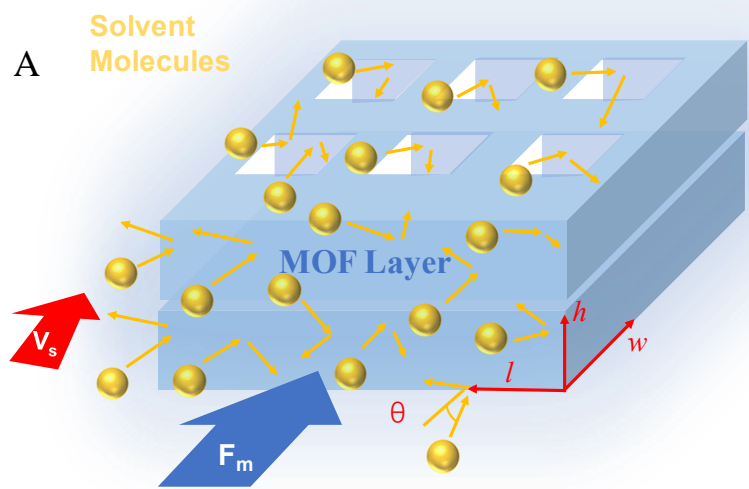

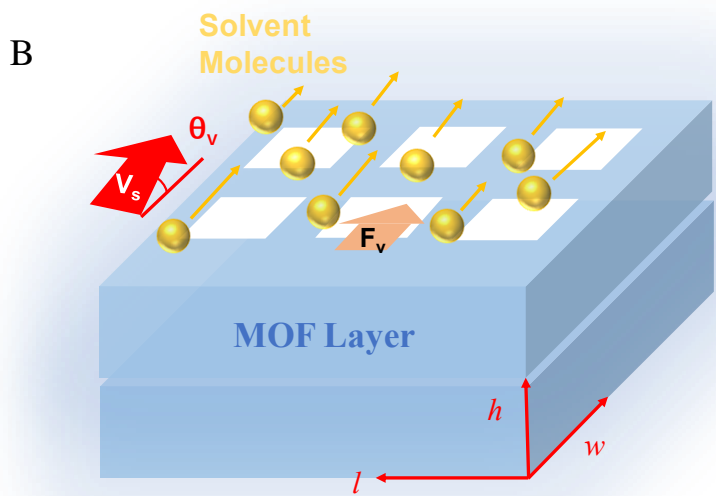

**Fig. S1.** The schematic illustration of the force in the exfoliation process, (A) $F_m$, (B) $F_v$. The golden balls represent the solvent molecules, the blue blocks represent MOF layer and locate in the $lw$ plane.

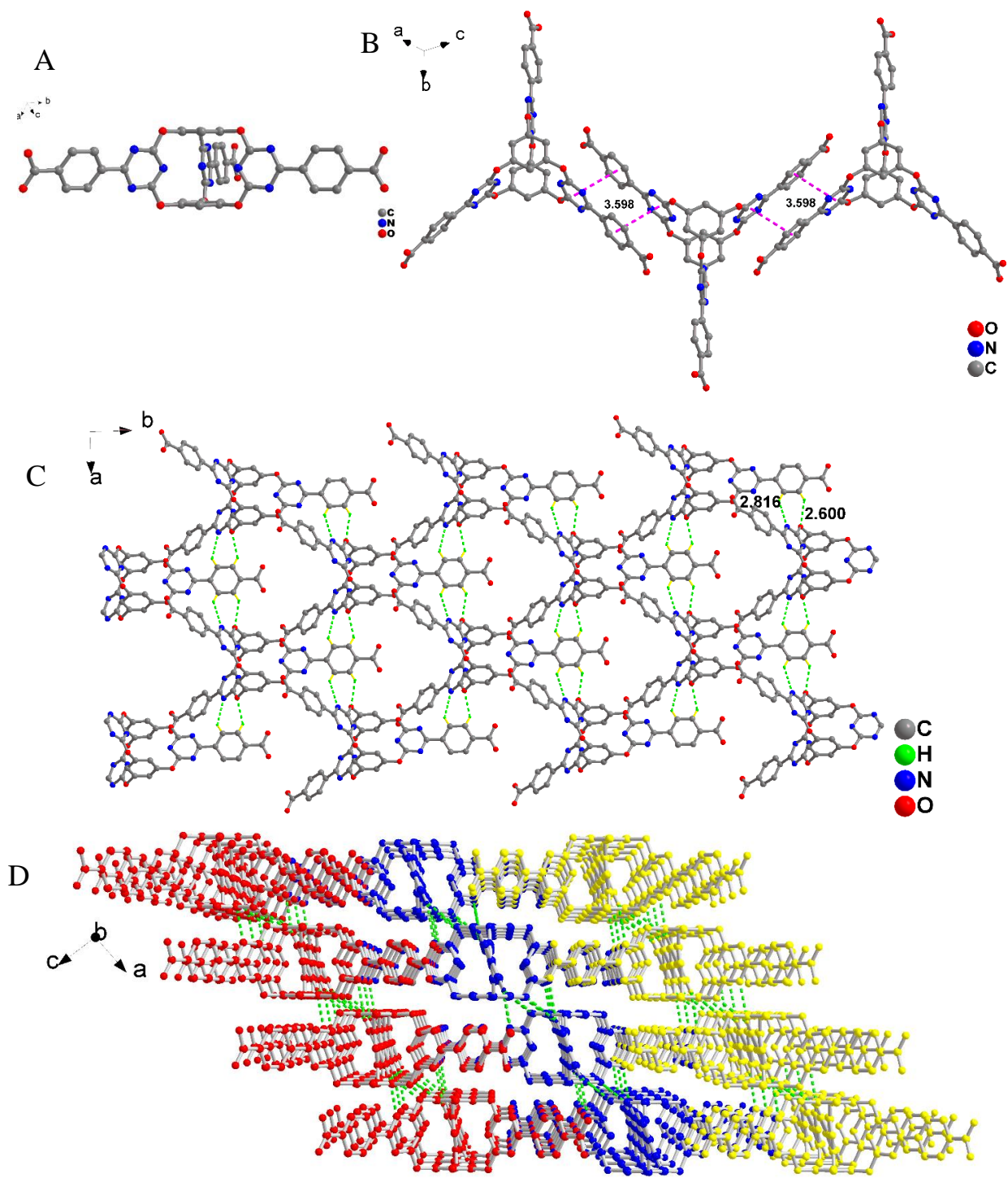

**Fig. S2.** Crystal structure of **BCTA** (A) Structure of single **BCTA** molecule. (B) The π -π stacking interaction between the **BCTA** chain in the layer. (C) The 2D layer structure connected via hydrogen bonds. (D) The 3D supramolecular framework constructed via hydrogen bonds and π -π stacking interaction.

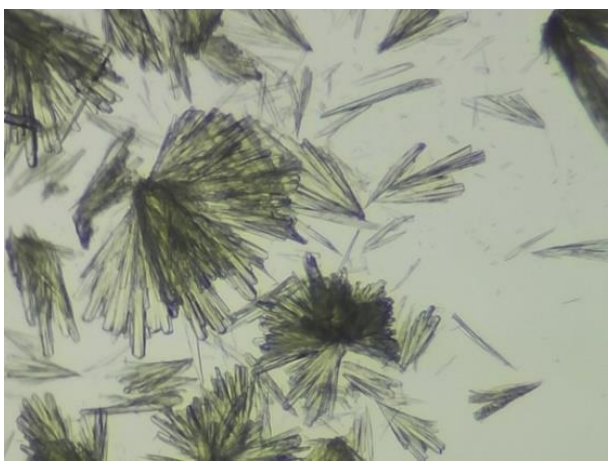

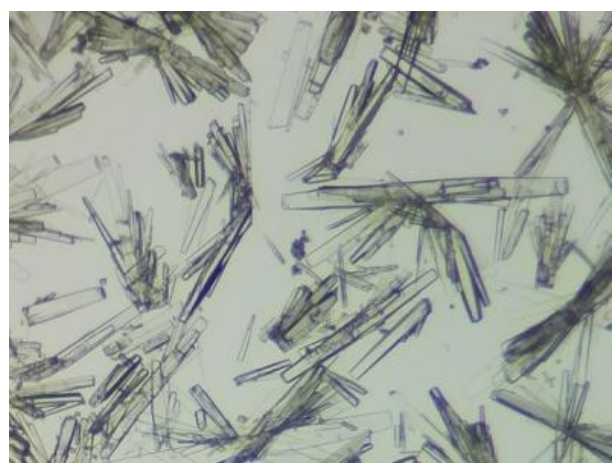

**Fig. S3.** Photos of **CSUMOF-1** crystals (A) and **CSUMOF-2** crystals (B).

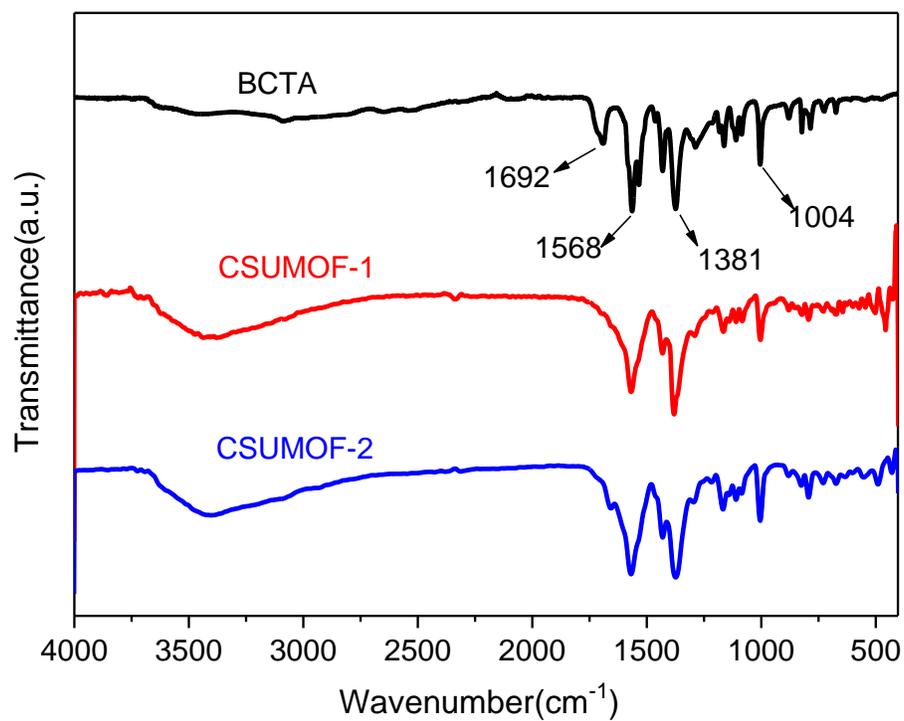

**Fig. S4.** Infrared spectra of **BCTA** (black), **CSUMOF-1** (red), **CSUMOF-2** (blue)

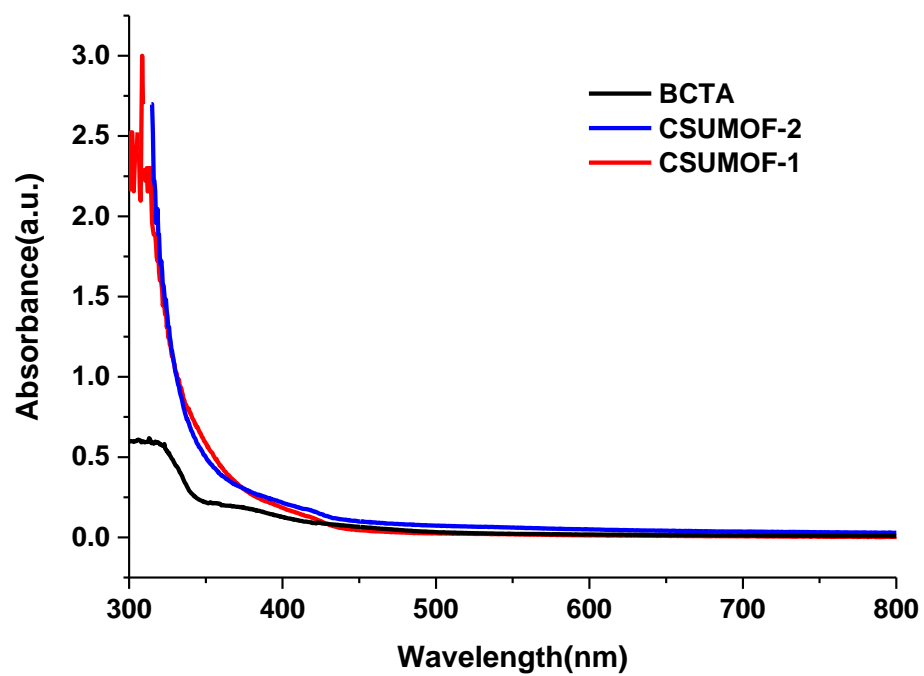

**Fig. S5.** UV-Vis spectra of **BCTA** (black), **CSUMOF-1** (red), **CSUMOF-2** (blue)

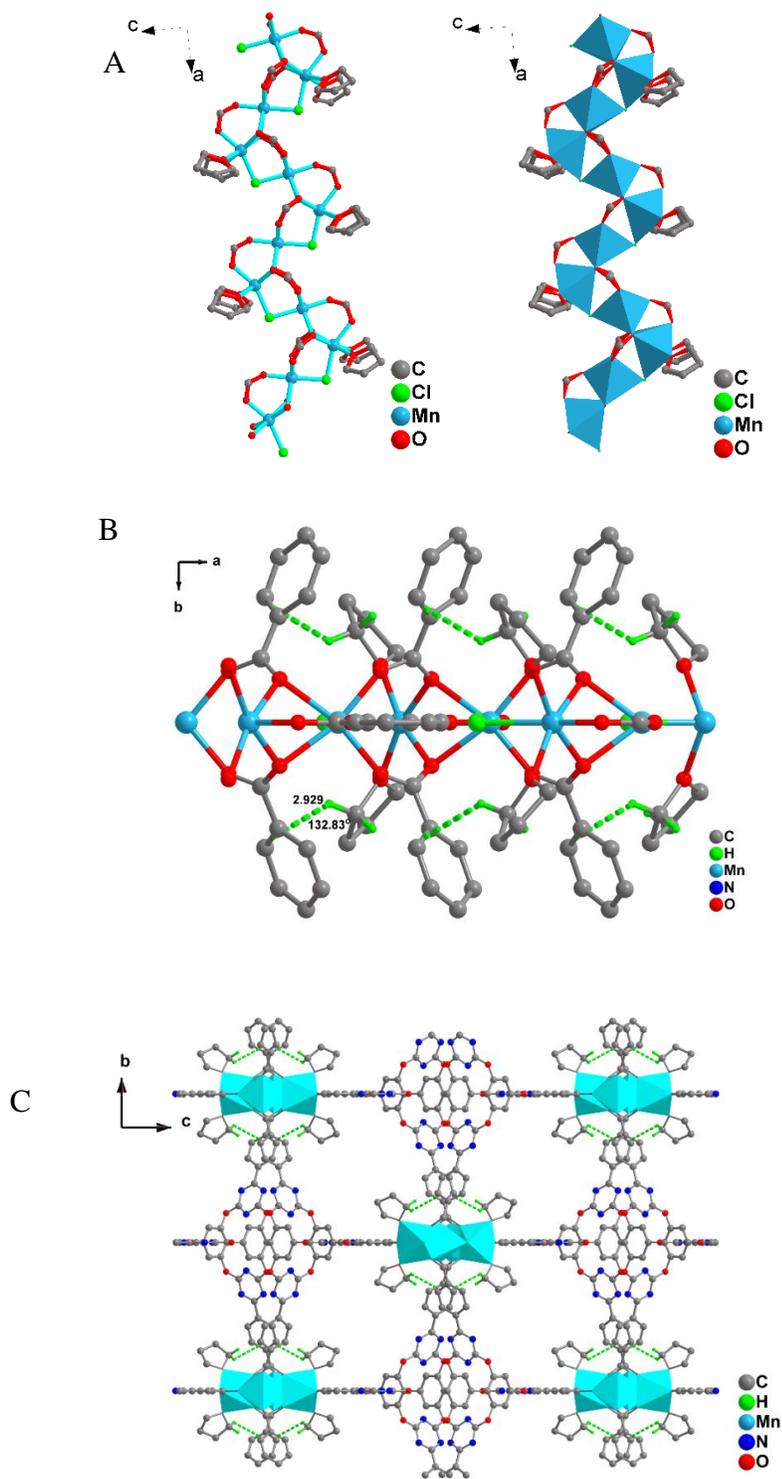

**Fig. S6.** (A) 1D sinusoidal [–Mn-Cl-Mn-O-]$_n$ inorganic chain in **CSUMOF-2,** view along *a* axis. (B) The C–H···π bonds between coordinated THF molecules and benzene rings in **CSUMOF-2**, view along [0, 0, 2] axis. (C) The C–H···π bonds between coordinated THF molecules and benzene rings in **CSUMOF-2**, view along *a* axis.

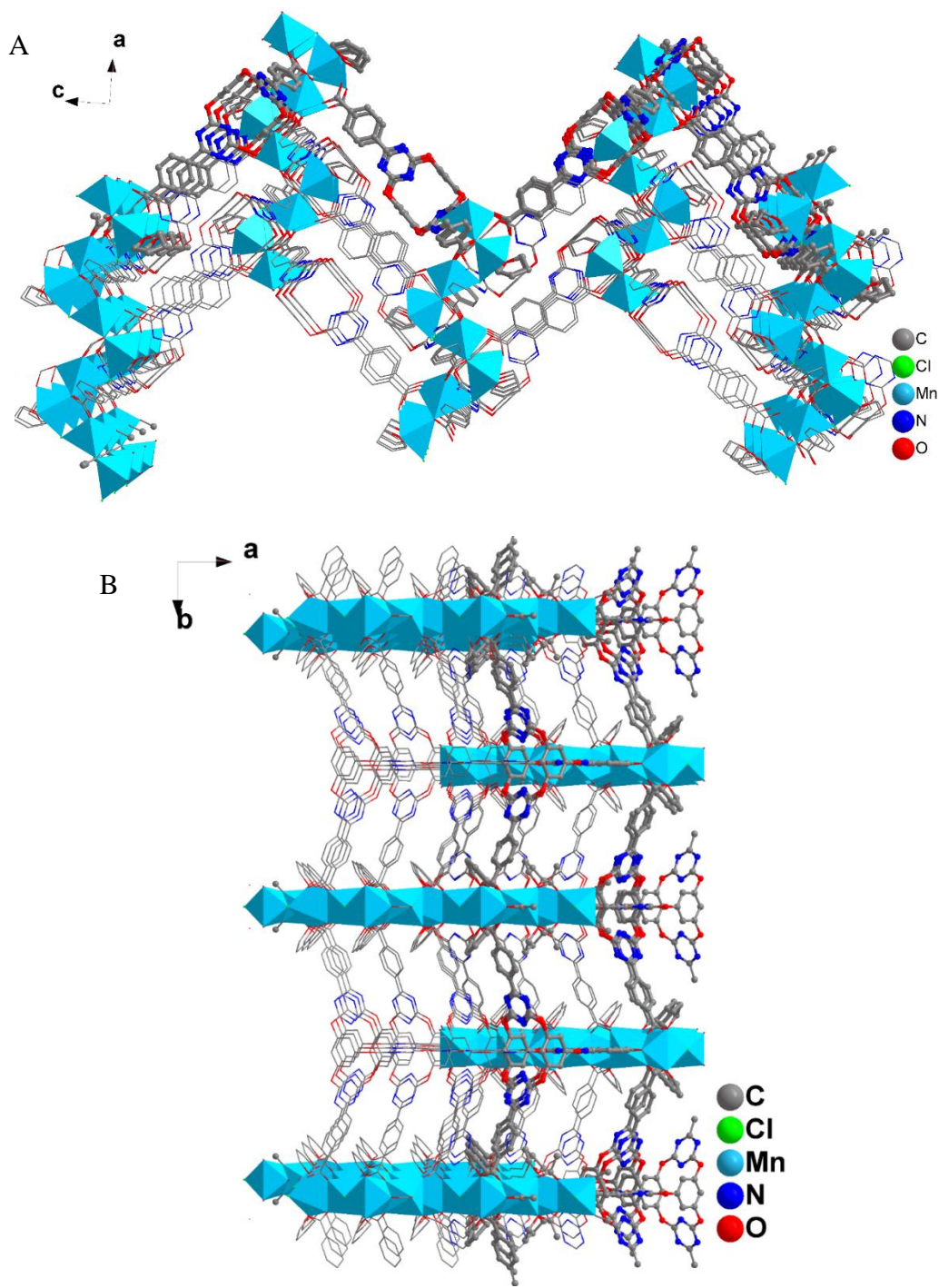

**Fig. S7.** (A) The Zigzag like layer structure of **CSUMOF-2**, view along [0, 2, 0] axis. (B) The 3D frame structure of **CSUMOF-2**, view along [0, 0, 2] axis.

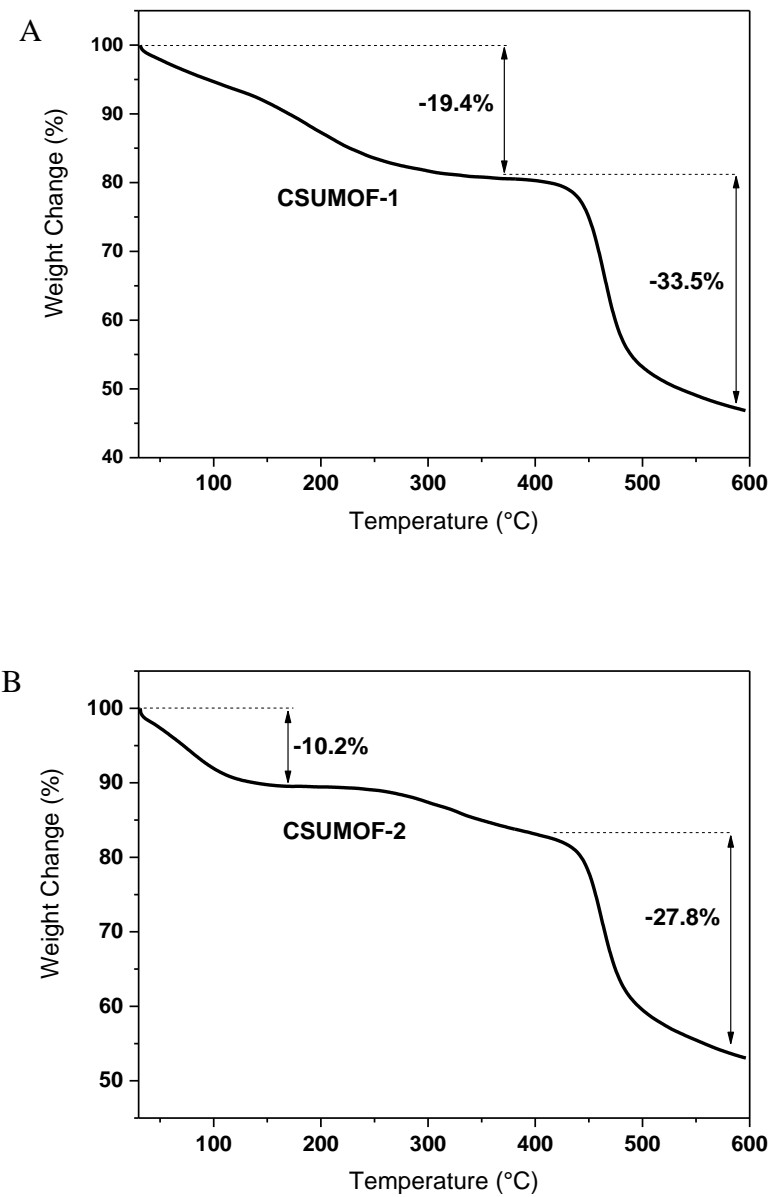

**Fig. S8.** TG analysis of **CSUMOF-1**(A) and **CSUMOF-2** (B)

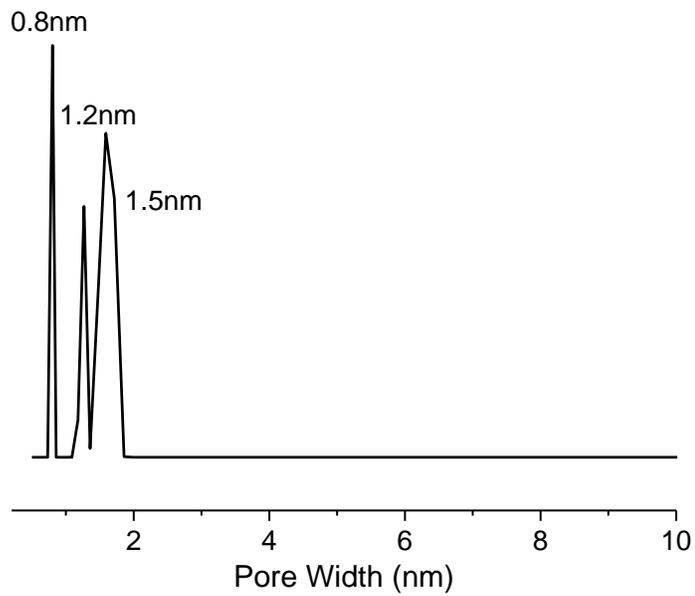

**Fig. S9.** Pore size distribution pattern of activated **CSUMOF-2** crystals.

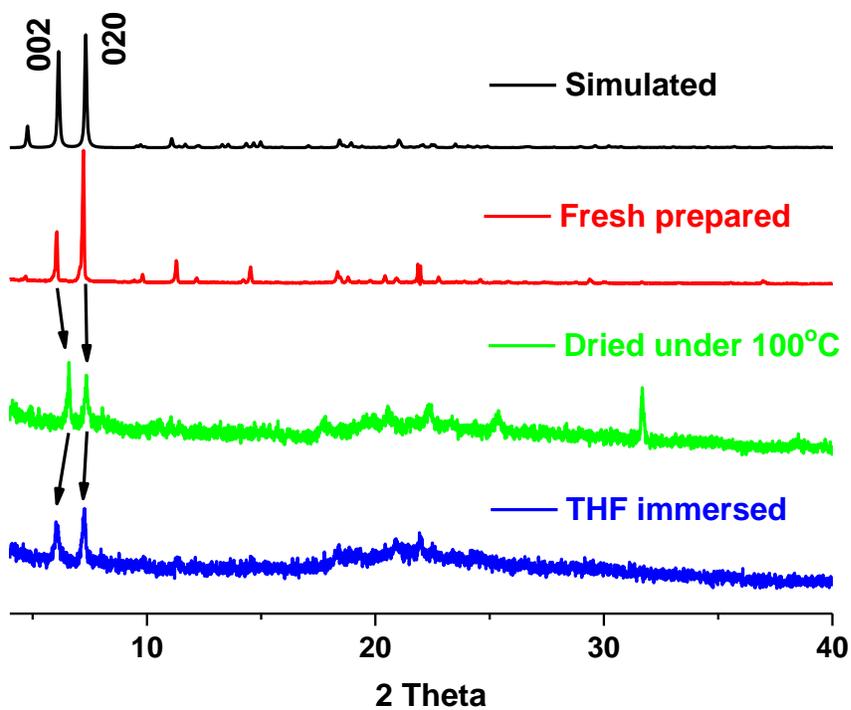

**Fig. S10.** XRD powder patterns of **CSUMOF-2** crystals. The simulated powder pattern based on single crystal data (black), fresh prepared (red), after drying at 100 °C for 3h (green), soaked in THF for 3h (blue).

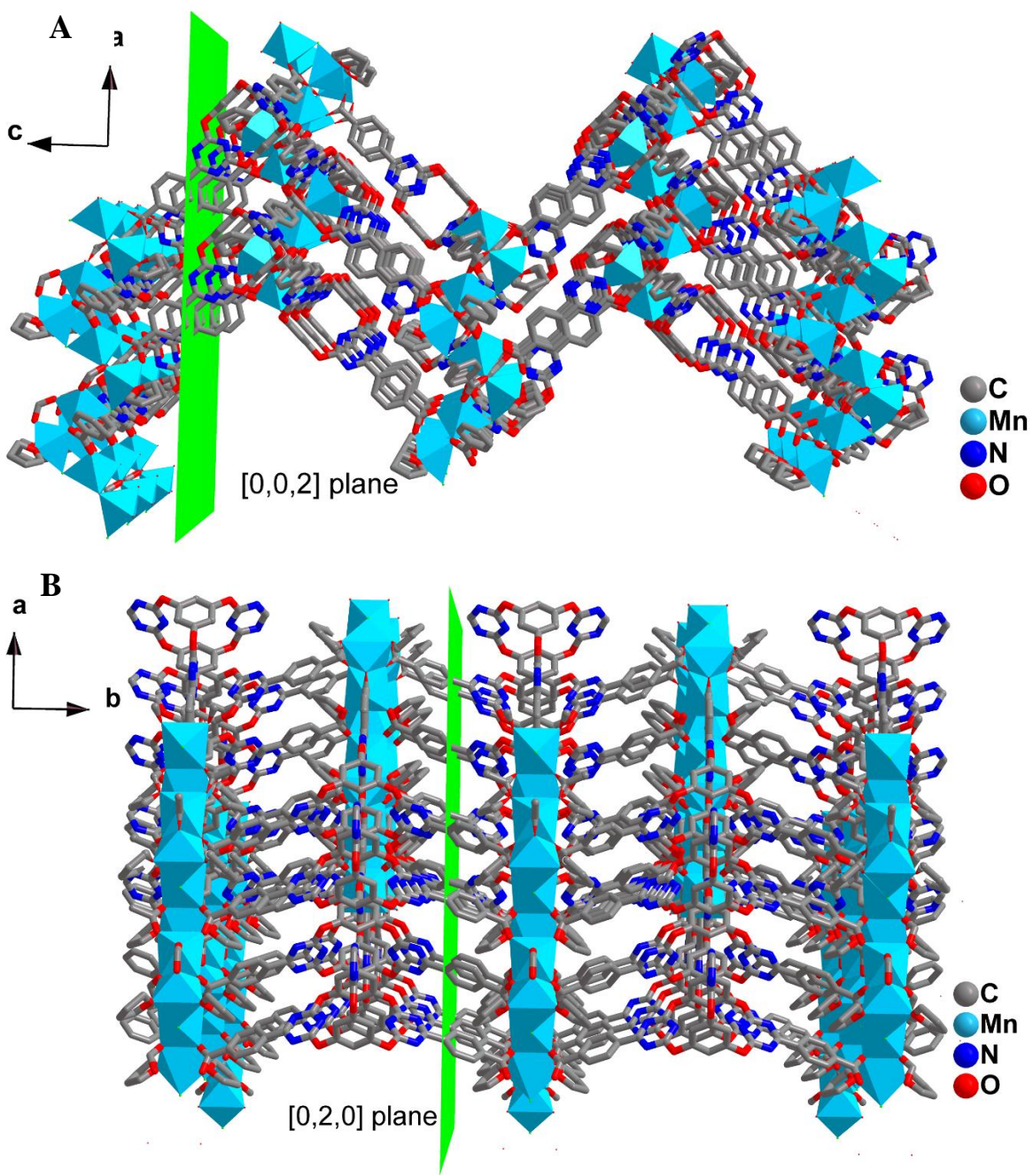

**Fig. S11.** (A) 002 and (B) 020 crystal planes of **CSUMOF-2**.

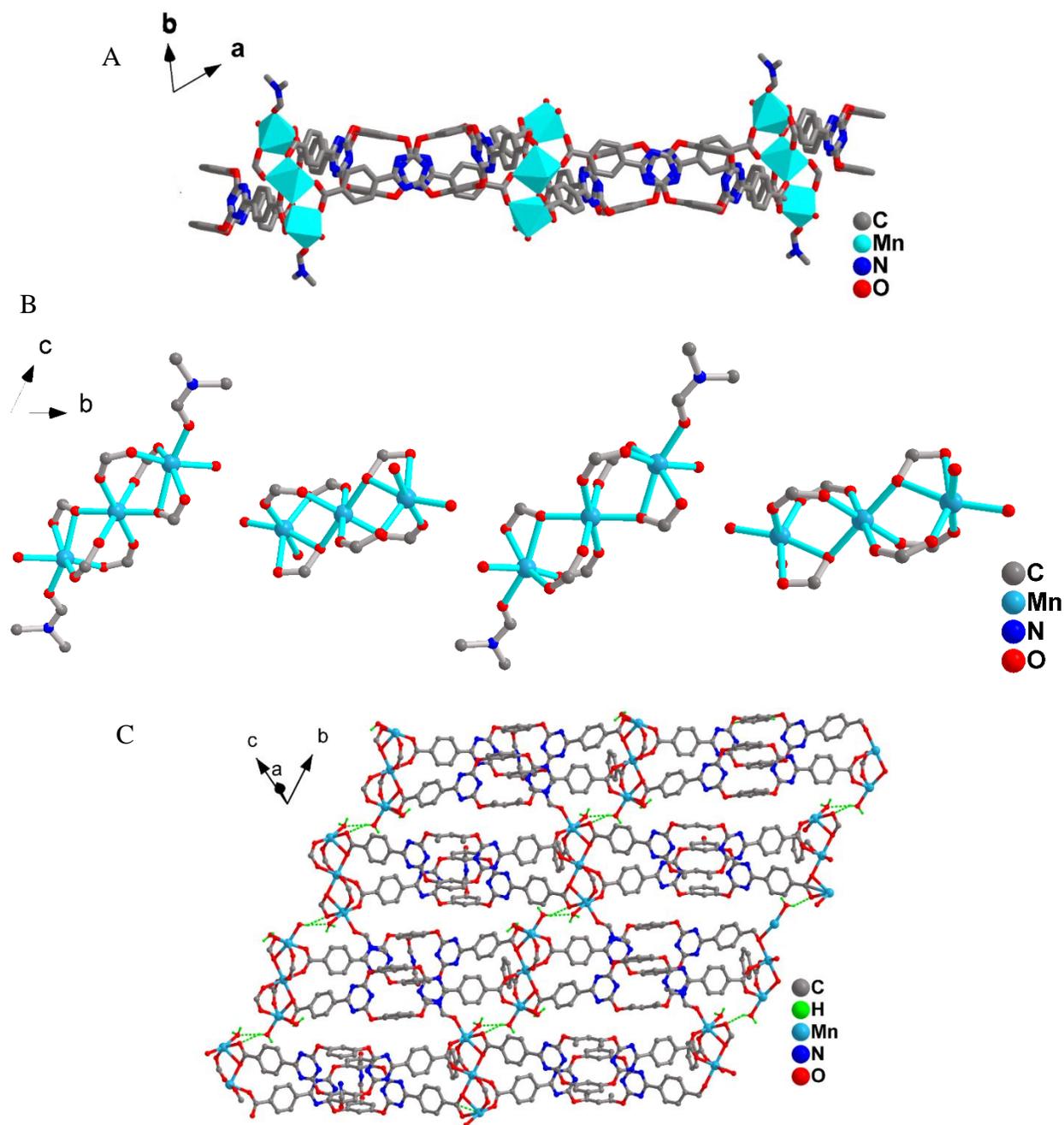

**Fig. S12.** (A) The presentation of one layer in **CSUMOF-1**, view along [0, -1, 1] axis. (B) The hydrogen bonds chain in **CSUMOF-1** composed by the two kinds of linear chain-like [$Mn_3(O_2C)_6$] clusters. (C) The three-dimensional neutral supramolecular framework constructed *via* hydrogen bonds in **CSUMOF-1**.

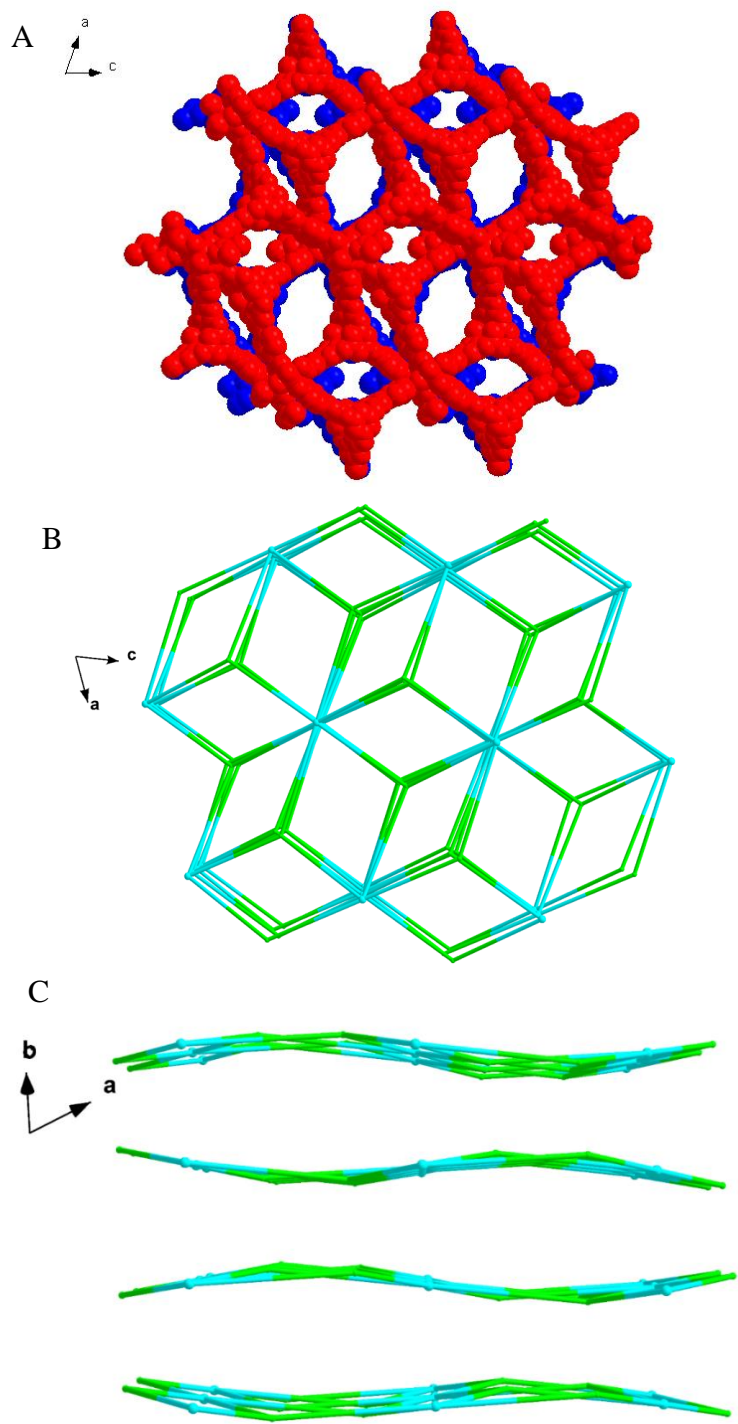

**Fig. S13.** (A) The three-dimensional framework with ellipsoid channels in **CSUMOF-1.** (B) The three-dimensional framework with regular channels in **CSUMOF-1.** (C) The ABAB stacking fashion with the 3,6-connected net nodes in **CSUMOF-1**, view along [0, -1, 1] axis.

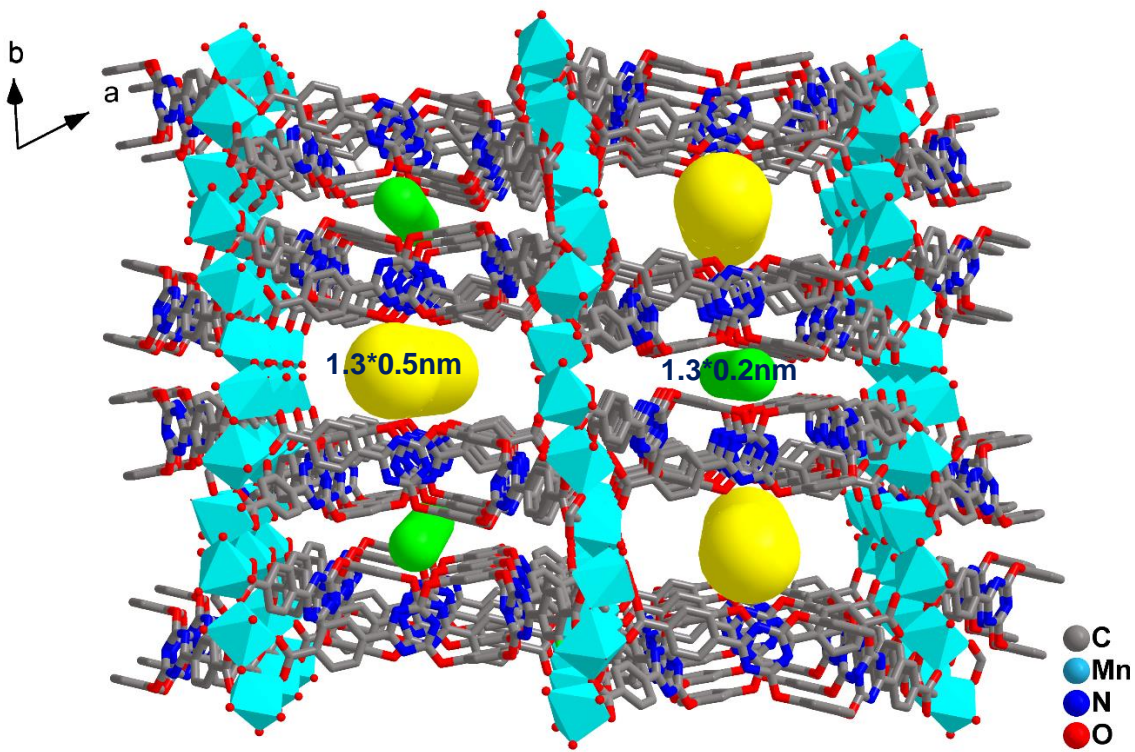

**Fig. S14.** Two types of channels with different size between the CSUMOF-1 layers.

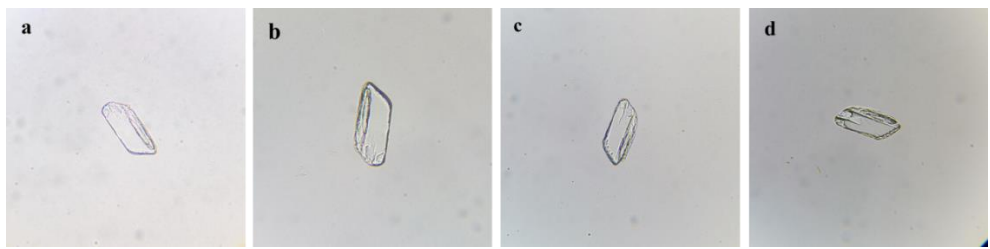

**Fig. S15.** The micrographs of one piece of **CSUMOF-1** crystal (a) as-prepared, (b) after washed by ethanol and dried, (c) after immersed in ethanol for 1 day, (d) after immersed in dichloromethane for 1 day.

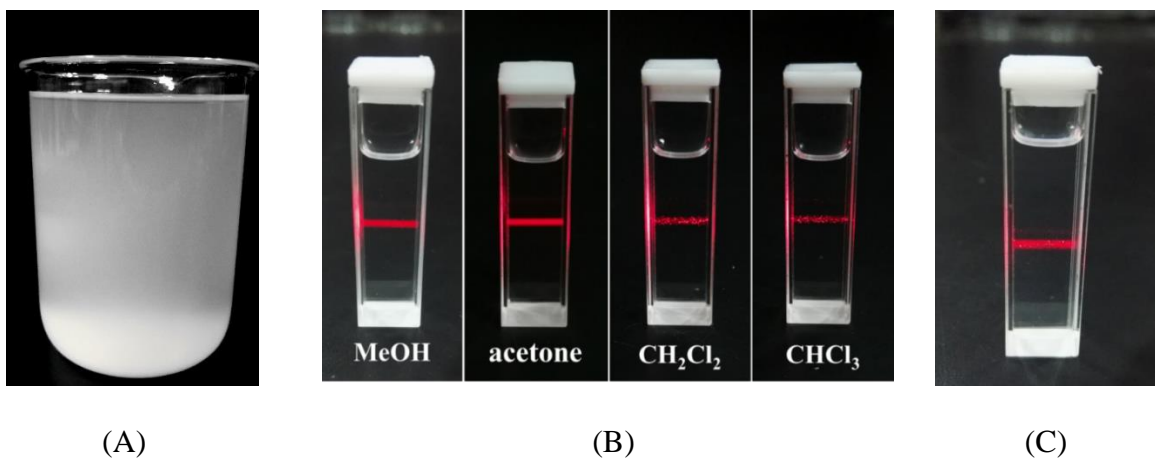

(A)                                    (B)                                  (C)

**Fig. S16.** (A) The high concentration suspension (0.2-0.4mg/ml) of **CSUMOF-1** prepared in large batch (500 ml). (B) Tyndall effect of the low concentration suspension of **CSUMOF-1** prepared in different solvents. (C) Tyndall effect of the suspension of **CSUMOF-2** in ethanol.

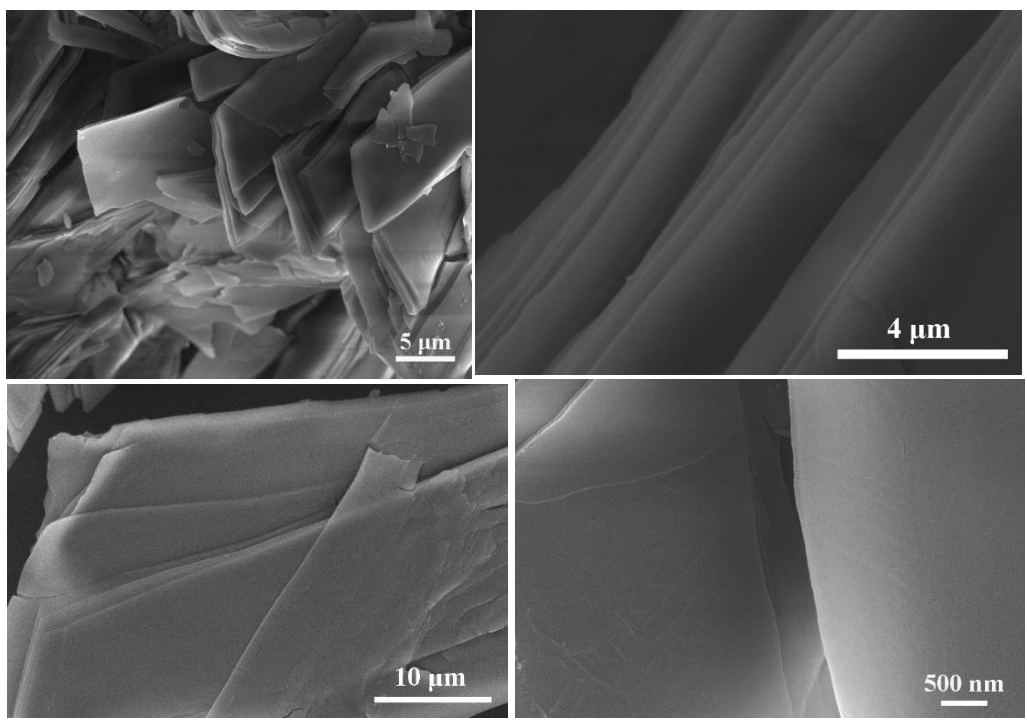

**Fig. S17.** The layered structure observed in the SEM images of **CSUMOF-1** crystal at different zoom levels.

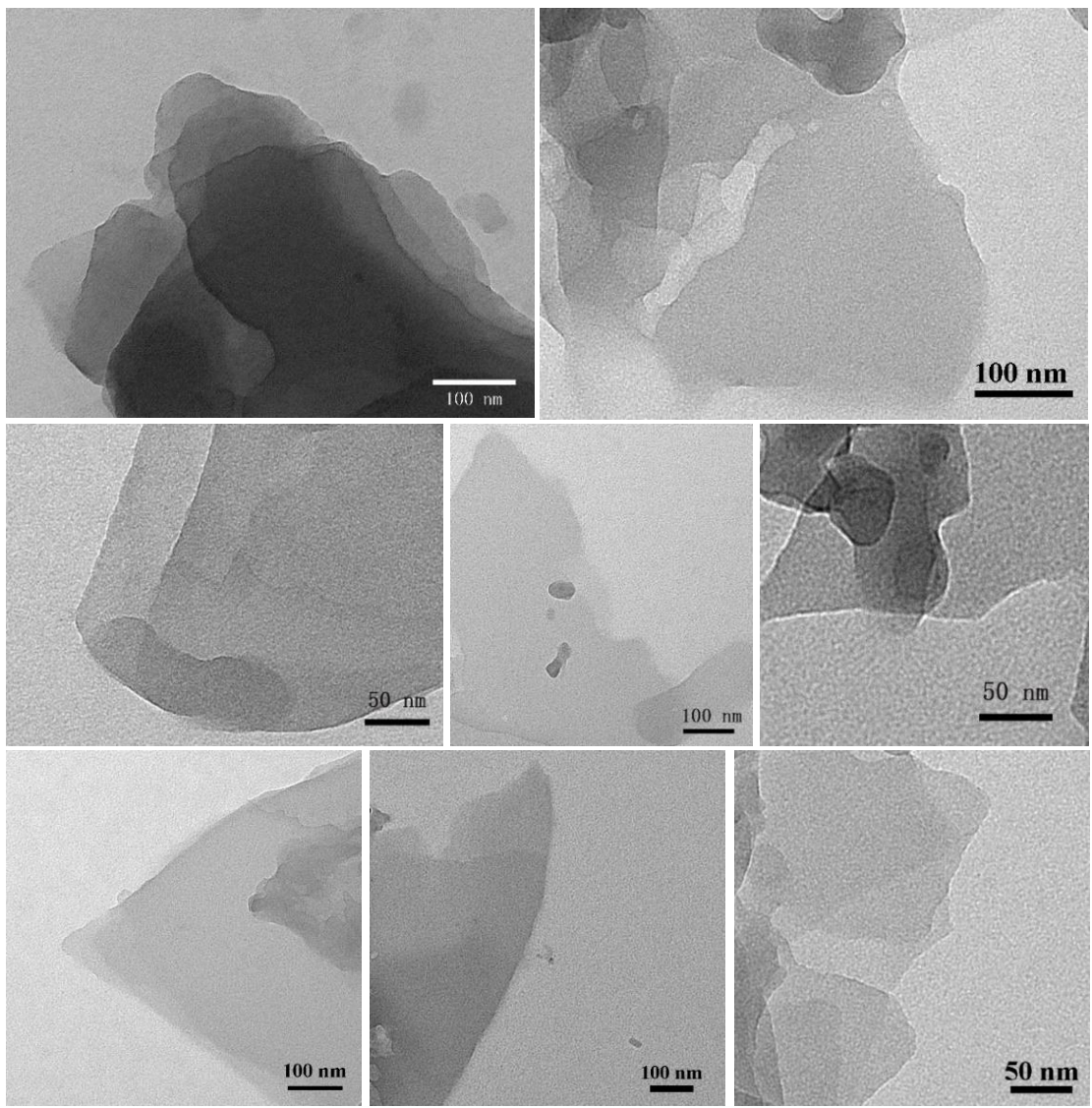

**Fig. S18.** TEM images of **CSUMOF-1** nanosheets.

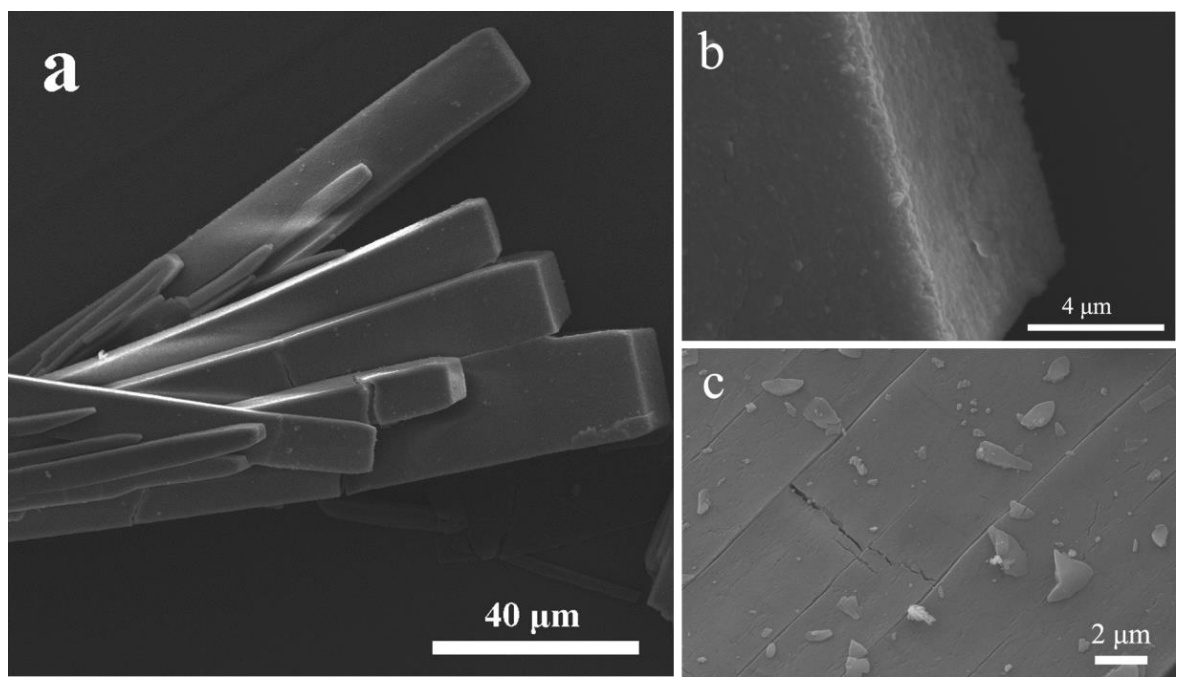

**Fig. S19.** SEM images of **CSUMOF-2** crystals at different zoom scale.

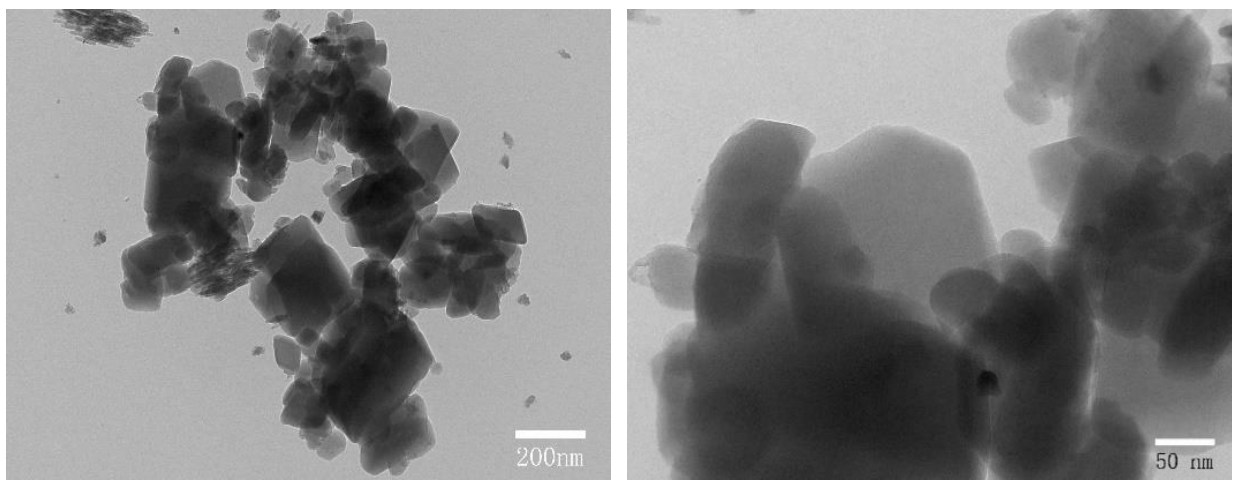

**Fig. S20.** SEM images of nano-particles in sonicated suspension of **CSUMOF-2**.

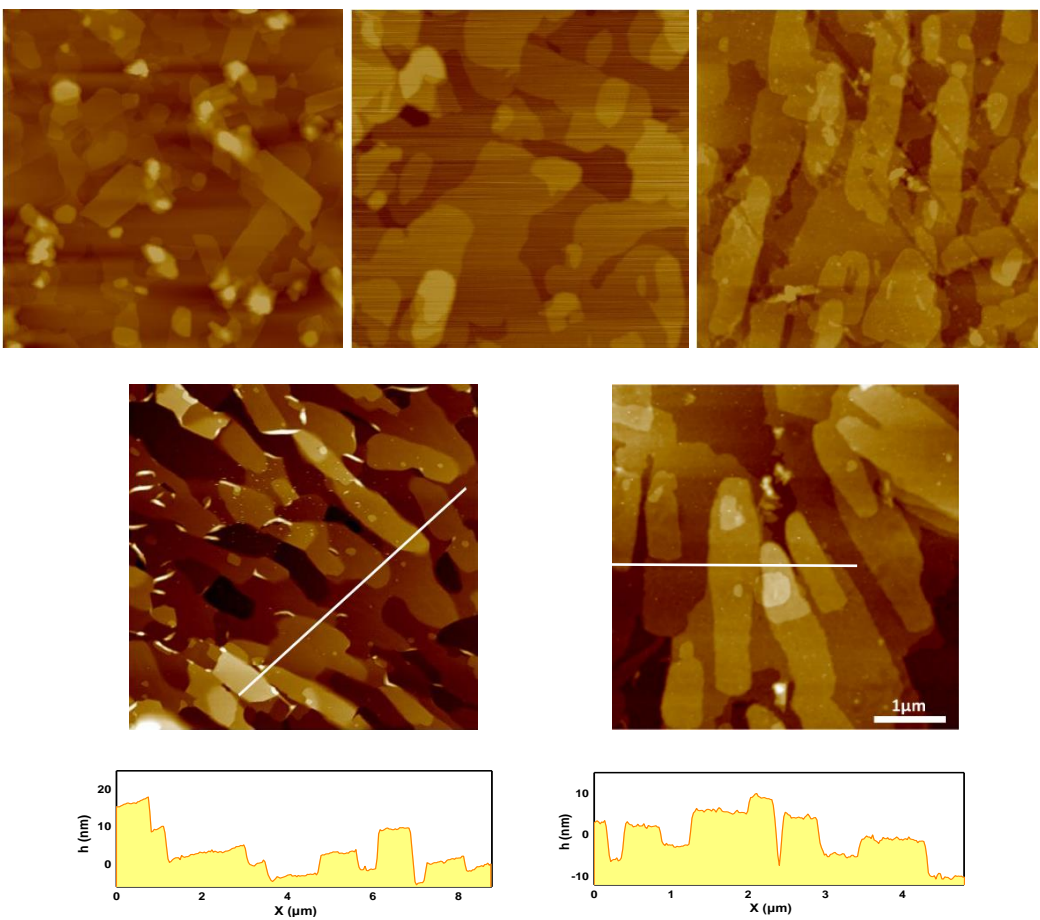

**Fig. S21.** AFM images of **CSUMOF-1** nanosheets suspension at high concentration.

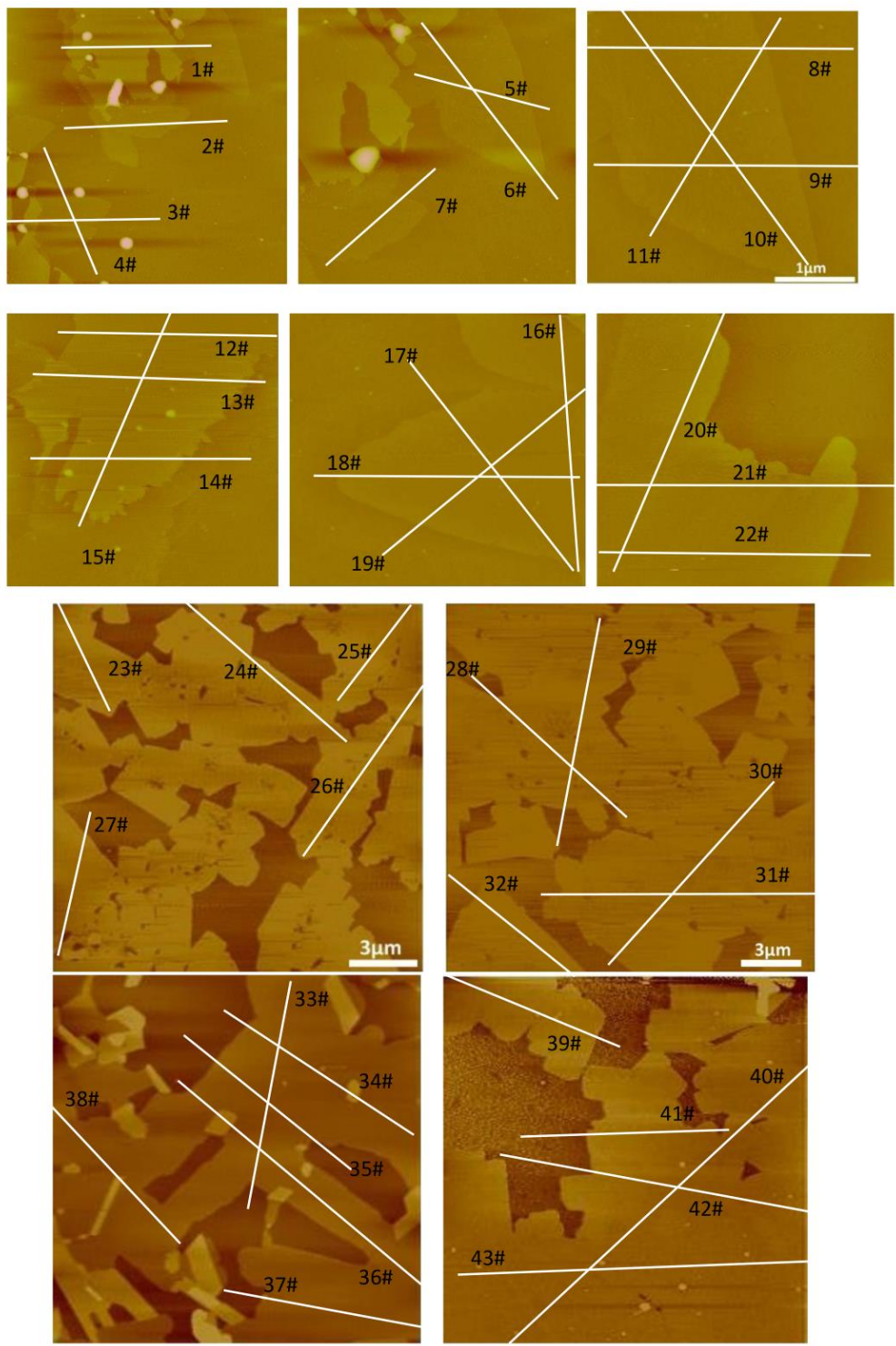

**Fig. S22.** AFM images of **CSUMOF-1** nanosheets suspension at low concentration.

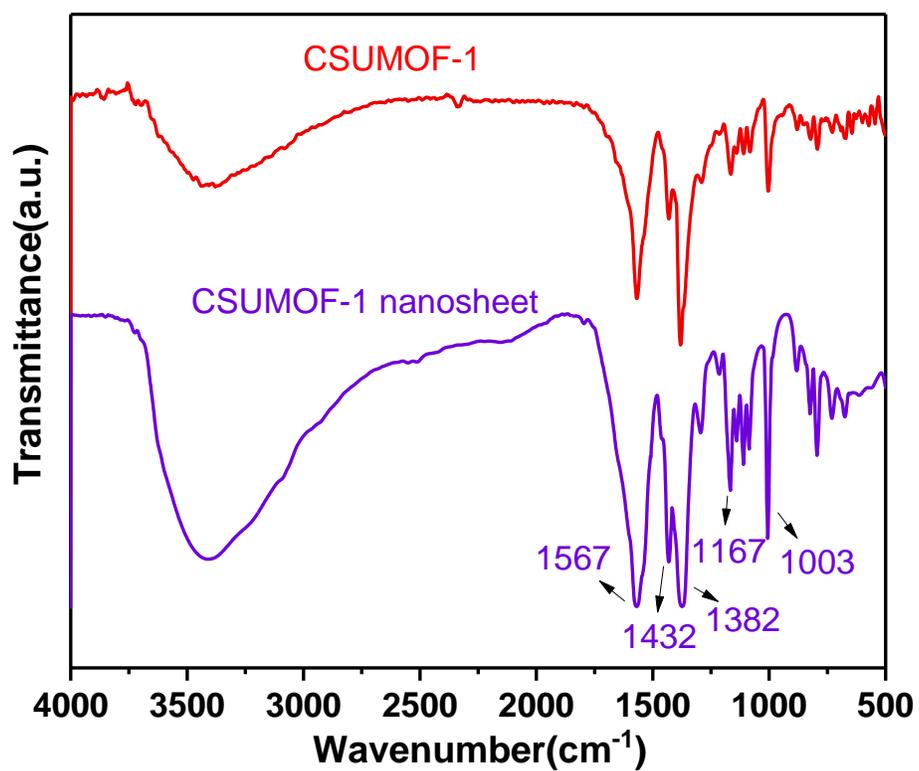

**Fig. S23.** Compare infrared spectra of **CSUMOF-1** (red) and **CSUMOF-1** nanosheet (purple)

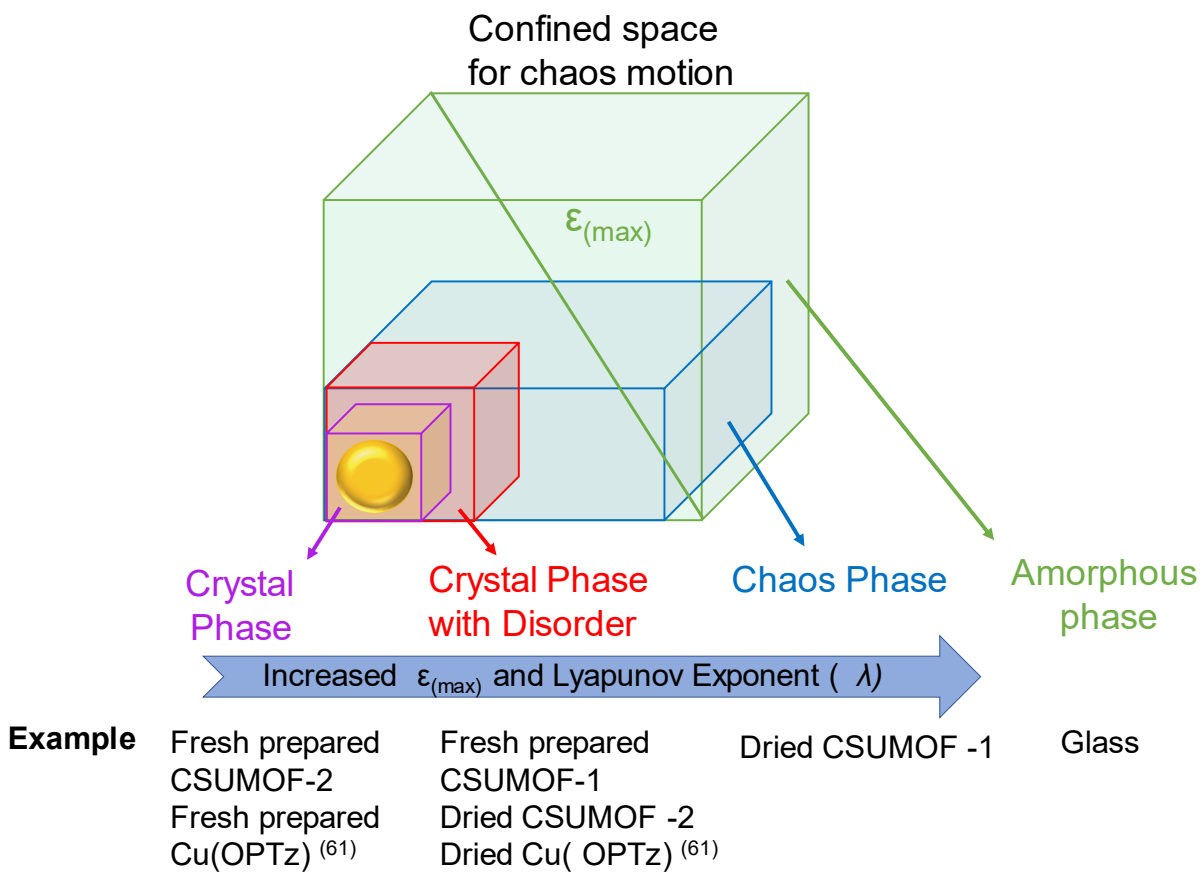

**Fig. S25.** The schematic illustration for the relationship between the confined space for chaos motion, $\varepsilon_{(max)}$, Lyapunov exponent and the corresponding material phase.